\documentclass[12pt]{article}
\usepackage{amsmath,amssymb}
\usepackage{wrapfig}
\usepackage{epsfig}
\usepackage{fancybox}
\usepackage{bm}
\usepackage{mathrsfs}
\usepackage{amsthm}
\usepackage{enumerate}
\usepackage{comment}
\usepackage{fourier}
\usepackage{multirow}
\allowdisplaybreaks[4]
\newcommand{{\Slash}}{0\!\!\!\!\!\big/}
\usepackage[usenames]{color}
\usepackage{setspace}
\usepackage{arydshln}

\begin{document}

\title{Generalized Heisenberg Dynamics Revisited}

\author{
Yoshiharu \textsc{Kawamura}\footnote{E-mail: haru@azusa.shinshu-u.ac.jp}\\
{\it Department of Physics, Shinshu University, }\\
{\it Matsumoto 390-8621, Japan}\\
}


\maketitle
\begin{abstract}
Taking as a model the fact that Heisenberg's matrix mechanics 
was derived from Hamiltonian mechanics using the correspondence principle, 
we explore a class of dynamical systems involving discrete variables, 
with Nambu mechanics as the starting point. 
Specifically, we reconstruct an extended version of matrix mechanics 
that describes dynamical systems possessing physical quantities 
expressed through generalized matrices. 
Furthermore, we reconfirm that a multiple commutator 
involving generalized matrices can serve as a discrete (quantized) version 
of the Nambu bracket or the Jacobian.
\end{abstract}

\section{Introduction}

A hundred years ago, Heisenberg formulated matrix mechanics~\cite{Heisenberg}, 
one of the foundational frameworks of quantum mechanics.
In its formulation, Bohr's correspondence principle played a guiding role. 
The correspondence principle states that, in a certain limit, 
physical quantities in quantum theory correspond to those in classical theory. 
In matrix mechanics (Heisenberg dynamics), 
physical quantities are represented by Hermitian matrices~\cite{B&J,BH&J}, 
and the theoretical structure exhibits similarities to 
Hamiltonian mechanics (Hamiltonian dynamics). 
For example, Heisenberg's equation of motion, 
which is the fundamental equation of matrix mechanics, 
takes the form of Hamilton's canonical equations
with the Poisson brackets replaced by commutators (divided by the imaginary unit
and the reduced Planck constant).
The existence of such similarities is considered to 
be one of the reasons why the correspondence principle functions effectively.

A few months after the advent of matrix mechanics, 
Schr\"{o}dinger proposed 
wave mechanics~\cite{Schrodinger1,Schrodinger2,Schrodinger3,Schrodinger4}, 
which is based on the Schr\"{o}dinger equation as its fundamental equation, 
and successfully reproduced the results of matrix mechanics. 
Subsequently, London, Jordan and Dirac demonstrated the equivalence of these two formulations 
through transformation theory~\cite{London,Jordan,Dirac}. 
At present, within the framework of quantum mechanics, 
matrix mechanics is understood as the theoretical formulation 
in the Heisenberg picture, in which the operator associated with energy 
(the Hamiltonian) is diagonalized, 
while wave mechanics is understood as the formulation 
in the Schr\"{o}dinger picture, in which the operator associated with position is diagonalized.

Today, quantum mechanics and its extension to field theory, 
known as quantum field theory (collectively referred to here as quantum theory), 
are widely utilized across various areas of physics and have achieved remarkable success. 
In light of such success, there naturally arises an interest in 
exploring the foundations of quantum theory, 
investigating the scope of its applicability, 
and examining the potential for its extension. 
Indeed, motivated by such considerations, 
attempts have been made to extend matrix mechanics, 
and its fundamental structure has been studied 
from an algebraic perspective~\cite{YK1,YK2,YK3,YK4}. 
This generalized mechanics is regarded as an extension of matrix mechanics, 
as it constructs physical quantities using an extended version of matrices, 
starting from generalized relations 
inspired by Bohr's frequency condition and Ritz's combination rule (see Figure 1).
In Refs.~\cite{YK1,YK2}, taking Heisenberg's matrix mechanics as a model, 
we constructed a dynamics based on generalized matrices with three indices, 
which we named cubic matrix mechanics. 
The equation of motion in this mechanics is formulated 
using a triple commutator, and since it involves only one counterpart of the Hamiltonian, 
it shares a structural similarity with Hamiltonian mechanics.

\begin{figure}[htbp]
\includegraphics[width=10cm,bb=0 0 600 540]{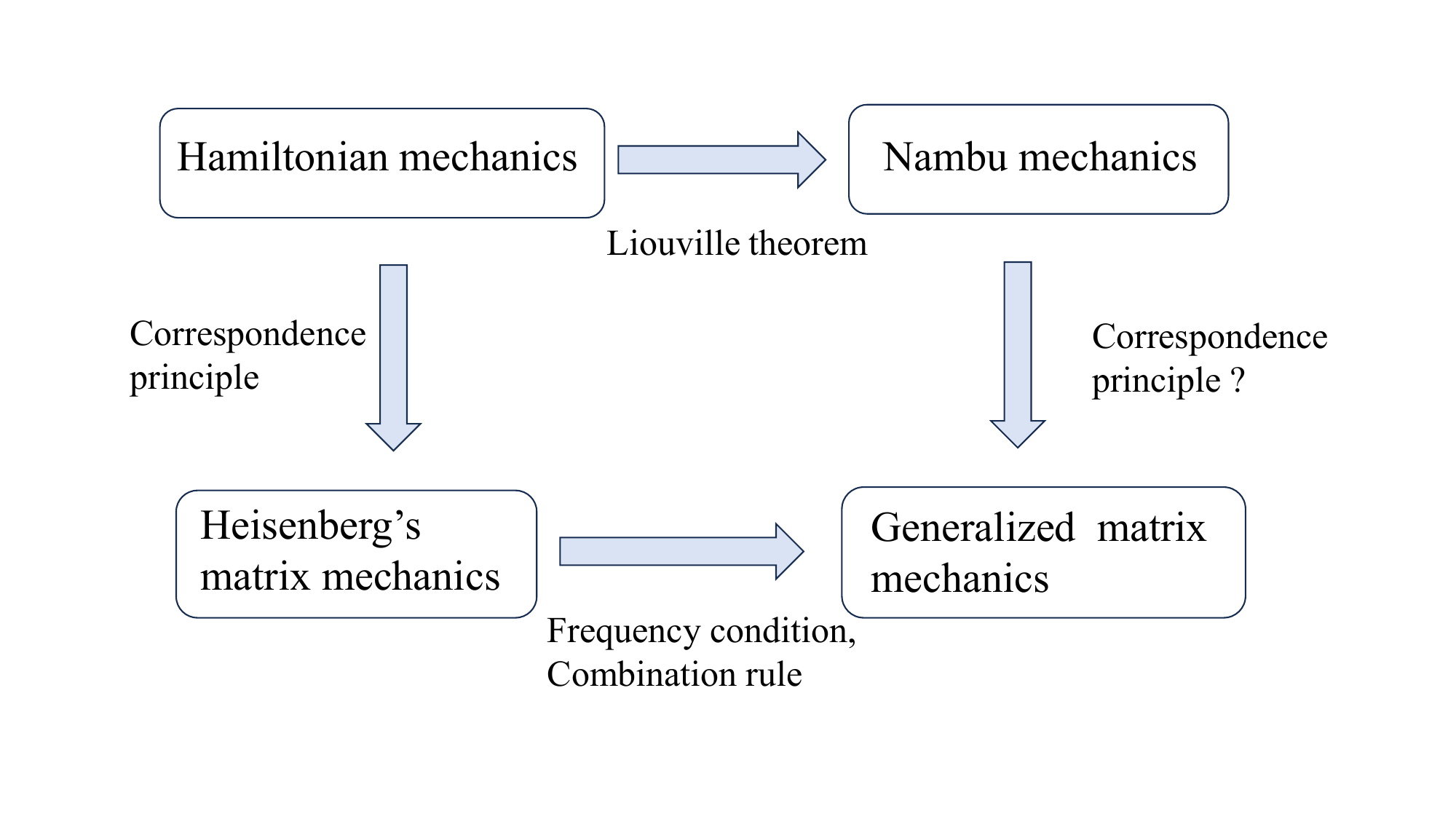}
\caption[F1]{Extensions of Hamiltonian mechanics and matrix mechanics.}
\label{F1}
\end{figure}

Furthermore, in Refs.~\cite{YK3,YK4}, the framework was extended to 
include several counterparts of the Hamiltonian, 
and hence this generalized matrix mechanics exhibits 
similarities to Nambu mechanics (Nambu dynamics) and can be regarded 
as a quantum version of Nambu mechanics. 
Here, Nambu mechanics is a generalization of Hamiltonian dynamics 
through the extension of phase space based on the Liouville theorem~\cite{Nambu} (see Figure 1). 
Specifically, in Ref.~\cite{YK3}, the relationship between Nambu mechanics 
based on canonical triplets and the dynamics involving extended matrices 
with three indices is discussed from the viewpoint of the correspondence principle. 
However, this analysis is somewhat ad hoc and remains insufficient. 
Moreover, it is unclear whether a similar correspondence holds 
in the general case of dynamics involving extended matrices with $n$ indices. 
In Ref.~\cite{YK4}, the theoretical formulation of dynamics 
based on extended matrices with $n$ indices was explored, 
but explicit expressions of frequency-related equations 
in terms of the eigenvalues of the Hamiltonians were not provided
in a definite form such that they possess those classical counterparts. 
In other words, further clarification of the theoretical structure of dynamics 
based on extended matrices with $n$ indices remains an open problem.

In this study, we address the question:
``What kind of theoretical structure emerges when the correspondence principle 
is applied to Nambu mechanics?''.
Answering this question is of considerable physical significance. 
This is because such an investigation can shed light on 
the `quantum' or discrete version of the Nambu bracket, 
namely the quantum analogue of the Jacobian, 
potentially leading to a formulation of quantum theory for extended objects\footnote{
See Refs.~\cite{H&M,Yoneya} for a comprehensive review of some aspects of Nambu mechanics
and its applications in string/M-theory.}. 
Furthermore, obtaining more concrete expressions for frequencies 
is expected to provide a clearer understanding of 
the underlying theoretical structure of an extended mechanics.

In this paper, we construct a framework of mechanics containing discrete variables, 
using the correspondence principle as the guiding principle 
and taking Nambu mechanics as the starting point. 
Specifically, we reconstruct an extended version of Heisenberg's matrix mechanics 
that describes dynamical systems 
whose physical quantities are represented by generalized matrices.
As a byproduct, we reconfirm that a multiple commutator 
involving generalized matrices can serve as 
the quantum or discrete analogue of the Nambu bracket, namely the quantum version of Jacobian.

The outline of this paper is as follows. 
In the next section, we review the structure of Hamiltonian dynamics 
and demonstrate how matrix mechanics can be 
derived from it using the correspondence principle. 
In section 3, we construct a dynamical framework 
involving extended matrices with three indices, 
starting from Nambu mechanics based on canonical triplets 
and applying the correspondence principle. 
In section 4, we develop a dynamical theory 
involving extended matrices with $n$ indices 
and study the properties of $n$-fold commutators 
defined on generalized matrices. 
In the final section, we present our conclusions and discussions. 
A detailed derivation of the equations presented in section 4 
is provided in the Appendix A.
It is shown that the generalized Heisenberg equation can be reduced to 
Heisenberg's equation of motion by taking a specific form of Hamiltonians
in the Appendix B.

\section{From Hamiltonian dynamics to Heisenberg dynamics}

\subsection{Hamiltonian dynamics}

First, we review the structure of Hamiltonian dynamics~\cite{Goldstein},
using a periodic motion described by the canonical variables $q=q(t)$ and $p=p(t)$.
These variables satisfy Hamilton's canonical equations of motion:
\begin{eqnarray}
\frac{dq}{dt}=\frac{\partial H}{\partial p},~~
\frac{dp}{dt}=-\frac{\partial H}{\partial q},
\label{Hc-eq}
\end{eqnarray}
where $H=H(q, p)$ is the Hamiltonian.
Using eq.~\eqref{Hc-eq}, we find that a general variable $A=A(q, p)$ satisfies the equation:
\begin{eqnarray}
\frac{dA}{dt}=\{A, H\}_{\rm PB},
\label{dA/dt}
\end{eqnarray}
where $\{A, H\}_{\rm PB}$ is the Poisson bracket defined by
\begin{eqnarray}
\{A, H\}_{\rm PB} \equiv \frac{\partial(A, H)}{\partial(q, p)}
= \frac{\partial A}{\partial q}\frac{\partial H}{\partial p}
 -  \frac{\partial H}{\partial q}\frac{\partial A}{\partial p}.
\label{PB}
\end{eqnarray}
From eqs.~\eqref{dA/dt} and \eqref{PB}, we see that $H$ conserves, 
i.e., $dH/dt = \{H, H\}_{\rm PB} = 0$.

The infinitesimal canonical transformation is given by
\begin{eqnarray}
\delta_{G}A=\varepsilon \{A, G\}_{\rm PB},
\label{deltaG}
\end{eqnarray}
where $\varepsilon$ is an infinitesimal constant
and $G=G(q, p)$ is a generating function.
Then, using the derivation rule:
\begin{eqnarray}
\frac{d}{dt} \{A, B\}_{\rm PB} = 
\left\{\frac{dA}{dt}, B\right\}_{\rm PB} + \left\{A, \frac{dB}{dt}\right\}_{\rm PB}
\label{D-rule}
\end{eqnarray}
and the Jacobi identity:
\begin{eqnarray}
\{\{A, G\}_{\rm PB}, H\}_{\rm PB}
+ \{\{G, H\}_{\rm PB}, A\}_{\rm PB} + \{\{H, A\}_{\rm PB}, G\}_{\rm PB} = 0,
\label{J-id}
\end{eqnarray}
it is shown that the equation is invariant under the canonical transformation such that
\begin{eqnarray}
\frac{dA'}{dt}=\{A', H\}_{\rm PB},
\label{dA'/dt}
\end{eqnarray}
where $A' = A + \delta_{G}A$.
For completeness, the derivation of eq.~\eqref{dA'/dt} is given as follows,
\begin{eqnarray}
\frac{dA'}{dt}&=&\frac{d}{dt}\left(A + \delta_{G}A\right) 
= \frac{d}{dt}\left(A + \varepsilon \{A, G\}_{\rm PB}\right)
\nonumber \\
&=& \frac{dA}{dt} + \varepsilon \left(\left\{\frac{dA}{dt}, G\right\}_{\rm PB} 
+ \left\{A, \frac{dG}{dt}\right\}_{\rm PB}\right)
\nonumber \\
&=&\{A, H\}_{\rm PB} + \varepsilon \left(\{\{A, H\}_{\rm PB}, G\}_{\rm PB}
+ \{A, \{G, H\}_{\rm PB}\}_{\rm PB}\right)
\nonumber \\
&=& \{A, H\}_{\rm PB} + \varepsilon \{\{A, G\}_{\rm PB}, H\}_{\rm PB}
= \{A + \varepsilon \{A, G\}_{\rm PB}, H\}_{\rm PB} 
\nonumber \\
&=& \{A + \delta_{G}A, H\}_{\rm PB} = \{A', H\}_{\rm PB}.
\label{dA'/dt-der}
\end{eqnarray}
From eqs.~\eqref{dA/dt} and \eqref{deltaG}, we find that the time evolution
of variables is realized by the canonical transformation
whose generation function is the Hamiltonian.

Let us carry out the canonical transformation $(q, p) \to (\theta, J)$,
in which $\theta$ becomes a cyclic coordinate,
using a generating function $W=W(q, J)$.
The canonical pairs are related to each other in the following way,
\begin{eqnarray}
pdq - Hdt = -\theta dJ - H' dt + dW,
\label{thetaJ}
\end{eqnarray}
where $H' = H'(J)$ is a new Hamiltonian.
We notice that $H'$ does not depend on $\theta$.
New canonical variables $\theta$ and $J$ satisfy Hamilton's canonical equations of motion:
\begin{eqnarray}
\frac{d\theta}{dt}=\frac{\partial H'}{\partial J},~~
\frac{dJ}{dt}=-\frac{\partial H'}{\partial \theta} = 0.
\label{dtheta/dt}
\end{eqnarray}

From eq.~\eqref{dtheta/dt},
we find that $J$ becomes a constant and then $\theta=\theta(t)$ is given by
\begin{eqnarray}
\theta(t) = \nu t +\theta_0,
\label{theta(t)}
\end{eqnarray}
where $\theta_0$ is a constant and $\nu$ is also a constant given by
\begin{eqnarray}
\nu = \frac{dE}{dJ}.
\label{nu}
\end{eqnarray}
In eq.~\eqref{nu}, $E=E(J)$ is the value of $H'$.

From eq.~\eqref{thetaJ}, $p$, $\theta$, $H$ and $H'$ are given by
\begin{eqnarray}
p = \frac{\partial W}{\partial q},~~ 
\theta = \frac{\partial W}{\partial J},~~
H = H' = E,
\label{ptheta}
\end{eqnarray}
respectively.

Let us clarify the physical meanings of $\theta$, $\nu$ and $J$.
We require that $\theta$ should satisfy $\displaystyle{\oint d\theta = 1}$ during one cycle,
and then $\nu$ is regarded as a frequency.
Furthermore, we find that $J$ is identified with the area on the phase space:
\begin{eqnarray}
J = \oint_{\rm C} pdq,
\label{J}
\end{eqnarray}
from the relation:
\begin{eqnarray}
\oint d\theta = \oint d\left(\frac{\partial W}{\partial J}\right) =
\frac{d}{dJ} \oint dW = \frac{d}{dJ} \oint_{\rm C} \frac{\partial W}{\partial q}dq =
\frac{d}{dJ} \oint_{\rm C} pdq = 1,
\label{J-der}
\end{eqnarray}
where we use eq.~\eqref{ptheta} and C stands for a closed trajectory on the phase space.
The variable $\theta$ and $J$ are called the angle variable 
and the action variable, respectively.
Inserting the momentum $p=p(q, E)$ derived from $H(q, p)=E$ 
into eq.~\eqref{J}, we see that $J$ becomes a function of $E$.
Differentiating $J=J(E)$ by $E$ and using 
$(\partial p/\partial E)(\partial H/\partial p)=1$
and $\displaystyle{{dq}/{dt} ={\partial H}/{\partial p}}$, we obtain the relation:
\begin{eqnarray}
\frac{dJ}{dE} 
= \oint_{\rm C} \frac{\partial p(q,E)}{\partial E} dq 
= \oint_{\rm C} \frac{1}{{\partial H}/{\partial p}} dq 
= \oint_{\rm C} \frac{1}{{dq}/{dt}} dq 
= \int_0^T dt = T,
\label{dJ/dE}
\end{eqnarray}
where $T$ is a period.
From eq.~\eqref{dJ/dE}, eq.~\eqref{nu} is rederived as
\begin{eqnarray}
\frac{dE(J)}{dJ} = \frac{1}{T} = \nu.
\label{nuE}
\end{eqnarray}
Using eq.~\eqref{dJ/dE} and $T=1/\nu$, $J$ is rewritten as
\begin{eqnarray}
J = \int {dJ} = \int \frac{dE}{\nu}.
\label{dE/nu}
\end{eqnarray}

In the classical electromagnetism,
electromagnetic waves are emitted from a rotating charged particle 
and the frequency of light $\nu_{\rm cl}$ follows the relation:
\begin{eqnarray}
\nu_{\rm cl} = \nu \Delta n = \frac{dE(J)}{dJ} \Delta n,
\label{nu-cl}
\end{eqnarray}
where $\nu$ is a number of rotation of the charged particle
and $\Delta n$ is a natural number.
It is known that the formula \eqref{nu-cl} does not hold microscopically
and it is modified at the quantum level.

\subsection{Heisenberg dynamics}

Let us construct Heisenberg dynamics from Hamiltonian dynamics,
taking the correspondence principle as a guiding principle.
To go into more detail about the correspondence principle,
it states that {\it results in classical physics should be
derived as limiting cases of those in quantum physics.
In concretely, there is a one-to-one correspondence
between quantum quantities and classical quantities,
and values of quantum ones and relations among quantum ones
approach values of classical ones and relations among classical ones in the limit
where a quantum number $n$ becomes large.}
This can be expressed as
\begin{eqnarray}
A_n \xrightarrow{n:{\rm large}} A_{\rm cl},~~
B_n \xrightarrow{n:{\rm large}} B_{\rm cl},~~~~
A_n = B_n \xrightarrow{n:{\rm large}} A_{\rm cl} = B_{\rm cl},
\label{An}
\end{eqnarray}
where $A_n$ and $B_n$ are quantum counterparts of 
classical quantities $A_{\rm cl}$ and $B_{\rm cl}$, respectively.

We assume that the quantum value of energy $E_n$ and the classical one $E(J)$ 
are related to each other as
\begin{eqnarray}
E_n = E(hn)
\label{EnJ}
\end{eqnarray}
through the Bohr-Sommerfeld quantization condition: 
\begin{eqnarray}
J = \oint_{\rm C} pdq = \iint dp \wedge dq = hn,
\label{BS}
\end{eqnarray}
where $h$ is the Planck constant and $n$ is a natural number. 

In this case, we obtain the following correspondence relation:
\begin{eqnarray}
&~& \frac{E_n - E_{n-\Delta n}}{h}
=\frac{E(hn) - E(h(n-\Delta n))}{h}
=\frac{E(hn) - E(hn-h\Delta n)}{h\Delta n}\Delta n
\nonumber\\
&~& ~~~~~~~~~~~~~~~~~~~~~~ \xrightarrow{n \gg \Delta n} 
\frac{1}{h}\frac{dE(hn)}{dn} \Delta n = \frac{dE(J)}{dJ} \Delta n = \nu \Delta n
= \nu_{\rm cl},
\label{nuCoPri}
\end{eqnarray}
and hence we realize that $(E_n - E_{n-\Delta n})/h$ must be the frequency of photon
in the quantum world.
In this way, we obtain Bohr's frequency condition:
\begin{eqnarray}
\nu_{m~\!\! n}=\frac{E_{m} - E_{n}}{h},
\label{nu-qu}
\end{eqnarray}
where $m$ and $n$ are positive integers.
Eq.~\eqref{nu-qu} is the quantum counterpart of eq.~\eqref{nu-cl},
and this formula with specific energy eigenvalues
explains experimental data of spectrum of light 
emitting from atoms very well.
For example, the energy eigenvalues are given by
$E_n = - {hcR}/{n^2}$ 
($R$ : the Rydberg constant, $c$ : the speed of light) for the hydrogen atom.

Next, let us derive an equation of motion for quantum quantities
based on the frequency condition \eqref{nu-qu}.
We consider a periodic motion with the period $T$
and suppose that quantities have two indices.
Then, the quantities $(A_1)_{m~\!\! n}(t)$ and $(A_2)_{m~\!\! n}(t)$ have a feature
that $(A_1)_{m~\!\! n}(t+T)=(A_1)_{m~\!\! n}(t)$ and $(A_2)_{m~\!\! n}(t+T)=(A_2)_{m~\!\! n}(t)$
and can be described as
\begin{eqnarray}
(A_1)_{m~\!\! n}(t)=(A_1)_{m~\!\! n}(0) e^{2\pi i \nu_{m~\!\! n}t},~~
(A_2)_{m~\!\! n}(t)=(A_2)_{m~\!\! n}(0) e^{2\pi i \nu_{m~\!\! n}t},
\label{Amn}
\end{eqnarray}
where $\nu_{m~\!\! n}$ follows the condition \eqref{nu-qu}
and is also written by $\nu_{m~\!\! n} = n_{m~\!\! n}/T$ ($n_{m~\!\! n}$ : an integer).

If we require that the product $(A_1 A_2)_{m~\!\! n}(t)$ 
should take the same form as eq.~\eqref{Amn}, i.e., 
$(A_1 A_2)_{m~\!\! n}(t)=(A_1 A_2)_{m~\!\! n}(0)e^{2\pi i \nu_{m~\!\! n}t}$,
$(A_1)_{m~\!\! n}(t)$ and $(A_2)_{m~\!\! n}(t)$ become components of matrices
whose product is defined by
\begin{align}
(A_1 A_2)_{m~\!\! n}(t) \equiv \sum_k (A_1)_{m~\!\! k}(t) (A_2)_{k~\!\! n}(t),
\label{AB}
\end{align}
where we use Ritz's combination rule:
\begin{eqnarray}
\nu_{m~\!\! n} = \nu_{m~\!\! k} + \nu_{k~\!\! n}.
\label{R-rule}
\end{eqnarray}
It is easy to see that the rule \eqref{R-rule} holds for
$\nu_{m~\!\! n}$ given by eq.~\eqref{nu-qu}.

Because the diagonal components of $(A_1)_{m~\!\! n}(t)$ and $(A_2)_{m~\!\! n}(t)$
do not depend on time
as seen from eqs.~\eqref{nu-qu} and \eqref{Amn}, 
any conserved quantity $C_{m~\!\! n}$ satisfying $dC_{m~\!\! n}/dt = 0$ can be written
as $C_{m~\!\! n} = c_m \delta_{m~\!\! n}$ with a constant $c_m$.
In a closed system, the Hamiltonian conserves, 
and hence its element $H_{m~\!\! n}$ takes the form:
\begin{eqnarray}
H_{m~\!\! n} = E_m \delta_{m~\!\! n},
\label{Hmn}
\end{eqnarray}
using the energy eigenvalue $E_m$.
In matrix mechanics, the task is to diagonalize the Hamiltonian, 
which is a function of matrices, in order to determine its eigenvalues.

Using eqs.~\eqref{nu-qu}, \eqref{Amn} and \eqref{Hmn}, we derive the equation:
\begin{align}
\frac{d}{dt} A_{m~\!\! n}(t) 
&= 2\pi i \nu_{m~\!\! n} A_{m~\!\! n}(t) 
= 2\pi i \frac{E_{m} - E_{n}}{h} A_{m~\!\! n}(t)
\nonumber \\
&= \frac{2\pi i}{h}
\left(\sum_{k} E_m \delta_{m~\!\! k} A_{k~\!\! n}(t) 
 - \sum_{k}A_{m~\!\! k}(t) E_k \delta_{k~\!\! n}\right)
\nonumber \\
&= \frac{2\pi i}{h}\sum_{k}
\left(H_{m~\!\! k} A_{k~\!\! n}(t) - A_{m~\!\! k}(t) H_{k~\!\! n}\right)
\nonumber \\
&= \frac{2\pi i}{h} [H, A(t)]_{m~\!\! n} 
= \frac{1}{i \hbar} [A(t), H]_{m~\!\! n},
\label{H-eq-derive}
\end{align}
where $\displaystyle{\hbar (\equiv {h}/({2\pi}))}$ is the reduced Planck constant
and $[A(t), H]_{m~\!\! n}$ is a commutator defined by
\begin{eqnarray}
[A(t), H]_{m~\!\! n} 
\equiv \sum_{k} \left(A_{m~\!\! k}(t) H_{k~\!\! n} - H_{m~\!\! k} A_{k~\!\! n}(t)\right).
\label{[A,H]}
\end{eqnarray}

In this way, we have arrived at Heisenberg's equation of motion:
\begin{eqnarray}
\frac{d}{dt} A_{m~\!\! n}(t) = \frac{1}{i \hbar} [A(t), H]_{m~\!\! n},
\label{H-eq}
\end{eqnarray}
as the quantum counterpart of eq.~\eqref{dA/dt}.

The infinitesimal unitary transformation is defined by
\begin{eqnarray}
\delta_{G}A_{m~\!\! n}=\varepsilon [A, G]_{m~\!\! n},
\label{deltaG-H}
\end{eqnarray}
where $\varepsilon$ is an infinitesimal constant
and $G_{m~\!\! n}$ is a generator.
Then, in a similar way to eqs.~\eqref{D-rule}--\eqref{dA'/dt-der}, using the derivation rule:
\begin{eqnarray}
\frac{d}{dt} [A, B]_{m~\!\! n} = 
\left[\frac{dA}{dt}, B\right]_{m~\!\! n} + \left[A, \frac{dB}{dt}\right]_{m~\!\! n}
\label{D-rule-com}
\end{eqnarray}
and the Jacobi identity:
\begin{eqnarray}
\left[[A, G], H\right]_{m~\!\! n} 
+ \left[[G, H], A\right]_{m~\!\! n} + \left[[H, A], G\right]_{m~\!\! n} = 0,
\label{J-id-com}
\end{eqnarray}
it is shown that the equation is invariant under the above transformation such that
\begin{eqnarray}
\frac{d}{dt}A'_{m~\!\! n}=[A', H]_{m~\!\! n},
\label{dA'/dt-H}
\end{eqnarray}
where $A'_{m~\!\! n} = A_{m~\!\! n} + \delta_{G}A_{m~\!\! n}$.
From eqs.~\eqref{H-eq} and \eqref{deltaG-H}, we find that the time evolution
of variables is generated by the unitary transformation
whose generator is the Hamiltonian.

\subsubsection{Fermionic harmonic oscillator}

We study a harmonic oscillator relating to fermionic variables as an example. 
They are represented by the $2 \times 2$ Hermitian matrices whose components are given by
\begin{eqnarray}
\xi_{m~\!\! n}(t) = \frac{1}{\sqrt{2}} |\varepsilon_{m~\!\! n}| 
e^{-i\omega \varepsilon_{m~\!\! n} t},~~
\eta_{m~\!\! n}(t) = -\frac{i}{\sqrt{2}} \varepsilon_{m~\!\! n}
e^{-i\omega \varepsilon_{m~\!\! n} t},
\label{xi}
\end{eqnarray}
where $m$ and $n$ run from $1$ to $2$, $\omega$ is the angular frequency,
$\varepsilon_{1~\!\! 2}=-\varepsilon_{2~\!\! 1}=1$ and
$\varepsilon_{1~\!\! 1}=\varepsilon_{2~\!\! 2}=0$. 
These variables satisfy the anti-commutation relations:
\begin{eqnarray}
\left\{\xi(t), \xi(t)\right\}_{m~\!\! n} = \delta_{m~\!\! n},~~
\left\{\eta(t), \eta(t)\right\}_{m~\!\! n} = \delta_{m~\!\! n},~~
\left\{\xi(t), \eta(t)\right\}_{m~\!\! n} = 0,
\label{antiCR}
\end{eqnarray}
where $\{A, B\}_{m~\!\! n} \equiv \sum_k(A_{m~\!\! k}B_{k~\!\! n} + B_{m~\!\! k}A_{k~\!\! n})$.

Fermionic variables become solutions of Heisenberg's equation of motion:
\begin{eqnarray}
\frac{d}{dt} \xi_{m~\!\! n}(t) = \frac{1}{i \hbar} [\xi(t), H]_{m~\!\! n},~~
\frac{d}{dt} \eta_{m~\!\! n}(t) = \frac{1}{i \hbar} [\eta(t), H]_{m~\!\! n},
\label{H-eq-xi}
\end{eqnarray}
where the Hamiltonian is given by
\begin{eqnarray}
H_{m~\!\! n} = i \hbar \omega \sum_{k=1}^{2} \xi_{m~\!\! k}(t) \eta_{k~\!\! n}(t)
= -\frac{\hbar\omega}{2}\sum_{k=1}^{2}\varepsilon_{m~\!\! k}~\! \delta_{m~\!\! n}. 
\label{H-xi}
\end{eqnarray}
Since $\xi_{m~\!\! n}(t)$ and $\eta_{m~\!\! n}(t)$ follow the equations:
\begin{eqnarray}
\frac{d}{dt} \xi_{m~\!\! n}(t) = \omega {\eta}_{m~\!\! n}(t),~~
\frac{d}{dt} \eta_{m~\!\! n}(t) = - \omega {\xi}_{m~\!\! n}(t),
\label{dxi}
\end{eqnarray}
they obey the equation for simple harmonic motion:
\begin{eqnarray}
\frac{d^2}{dt^2} \xi_{m~\!\! n}(t) = - \omega^2 {\xi}_{m~\!\! n}(t),~~
\frac{d^2}{dt^2} \eta_{m~\!\! n}(t) = - \omega^2 {\eta}_{m~\!\! n}(t).
\label{d2xi}
\end{eqnarray}

The same system is described by the representation matrices of the ladder operators:
\begin{eqnarray}
&~& C_{m~\!\! n}(t) \equiv \frac{1}{\sqrt{2}}\left(\xi_{m~\!\! n}(t) 
+ i \eta_{m~\!\! n}(t)\right)
= \frac{1}{2} \left(|\varepsilon_{m~\!\! n}| + \varepsilon_{m~\!\! n}\right)e^{-i\omega t},~~
\label{C}\\
&~& C^{\dagger}_{m~\!\! n}(t) \equiv \frac{1}{\sqrt{2}}\left(\xi_{m~\!\! n}(t) 
- i \eta_{m~\!\! n}(t)\right)
= \frac{1}{2} \left(|\varepsilon_{m~\!\! n}| - \varepsilon_{m~\!\! n}\right)e^{i\omega t}.
\label{Cdag}
\end{eqnarray}
These variables obey the anti-commutation relations:
\begin{eqnarray}
\{C(t), C^{\dagger}(t)\}_{m~\!\! n} = \delta_{m~\!\! n},~~
\{C(t), C(t)\}_{m~\!\! n} = 0,~~
\{C^{\dagger}(t), C^{\dagger}(t)\}_{m~\!\! n} = 0,
\label{antiCR-C}
\end{eqnarray}
and satisfy Heisenberg's equation of motion:
\begin{eqnarray}
\frac{d}{dt} C_{m~\!\! n}(t) = \frac{1}{i \hbar} [C(t), H]_{m~\!\! n},~~
\frac{d}{dt} C^{\dagger}_{m~\!\! n}(t) = \frac{1}{i \hbar} [C^{\dagger}(t), H]_{m~\!\! n},
\label{H-eq-C}
\end{eqnarray}
where the Hamiltonian is written by
\begin{eqnarray}
H_{m~\!\! n} = \frac{1}{2}\hbar \omega [C^{\dagger}(t), C(t)]_{m~\!\! n}
= \hbar \omega \left(\sum_{k=1}^{2} C^{\dagger}_{m~\!\! k}(t) C_{k~\!\! n}(t)
- \frac{1}{2}\delta_{m~\!\! n}\right). 
\label{H-C}
\end{eqnarray}
Since $C_{m~\!\! n}(t)$ and $C^{\dagger}_{m~\!\! n}(t)$ follow the equations:
\begin{eqnarray}
\frac{d}{dt} C_{m~\!\! n}(t) = -i\omega C_{m~\!\! n}(t),~~
\frac{d}{dt} C^{\dagger}_{m~\!\! n}(t) = i \omega C^{\dagger}_{m~\!\! n}(t),
\label{dC}
\end{eqnarray}
they also obey the equation for simple harmonic motion.

\section{From Nambu dynamics to generalized Heisenberg dynamics}

\subsection{Nambu dynamics}

First, we review the structure of Nambu dynamics~\cite{Nambu},
using a periodic motion described by the canonical triplet $x=x(t)$, $y=y(t)$ and $z=z(t)$.
These variables satisfy a generalization of Hamilton's canonical equations of motion
called the Nambu equation:
\begin{eqnarray}
\frac{dx}{dt}=\frac{\partial (H_1, H_2)}{\partial (y, z)},~~
\frac{dy}{dt}=\frac{\partial (H_1, H_2)}{\partial (z, x)},~~
\frac{dz}{dt}=\frac{\partial (H_1, H_2)}{\partial (x, y)},
\label{N-eq}
\end{eqnarray}
where $H_1=H_1(x, y, z)$ and $H_2=H_2(x, y, z)$ are the Hamiltonians.
Using eq.~\eqref{N-eq}, we find that a general variable $A=A(x, y, z)$ satisfies the equation:
\begin{eqnarray}
\frac{dA}{dt}=\{A, H_1, H_2\}_{\rm NB},
\label{dA/dt-N}
\end{eqnarray}
where $\{A, H_1, H_2\}_{\rm NB}$ is the Nambu bracket defined by
\begin{align}
\{A, H_1, H_2\}_{\rm NB} &\equiv \frac{\partial(A, H_1, H_2)}{\partial(x, y, z)}
\nonumber \\
&= \frac{\partial A}{\partial x}\frac{\partial H_1}{\partial y}\frac{\partial H_2}{\partial z}
+ \frac{\partial H_1}{\partial x}\frac{\partial H_2}{\partial y}\frac{\partial A}{\partial z}
+ \frac{\partial H_2}{\partial x}\frac{\partial A}{\partial y}\frac{\partial H_1}{\partial z}
\nonumber \\
&~~~ 
- \frac{\partial H_2}{\partial x}\frac{\partial H_1}{\partial y}\frac{\partial A}{\partial z}
- \frac{\partial H_1}{\partial x}\frac{\partial A}{\partial y}\frac{\partial H_2}{\partial z}
- \frac{\partial A}{\partial x}\frac{\partial H_2}{\partial y}\frac{\partial H_1}{\partial z}.
\label{NB}
\end{align}
From eqs.~\eqref{dA/dt-N} and \eqref{NB}, we see that $H_1$ and $H_2$ conserve, i.e.,
$dH_1/dt = \{H_1, H_1, H_2\}_{\rm NB} = 0$ 
and $dH_2/dt = \{H_2, H_1, H_2\}_{\rm NB} = 0$.

The infinitesimal generalized canonical transformation is defined by
\begin{eqnarray}
\delta_{G}A=\varepsilon \{A, G_1, G_2\}_{\rm NB},
\label{deltaG-N}
\end{eqnarray}
where $\varepsilon$ is an infinitesimal constant
and $G_1=G_1(x, y, z)$ and $G_2=G_2(x, y, z)$ are the generating functions.
Then, in a similar way to eqs.~\eqref{D-rule}--\eqref{dA'/dt-der}, 
using the derivation rule:
\begin{eqnarray}
\frac{d}{dt} \{A, B, C\}_{\rm NB} = 
\left\{\frac{dA}{dt}, B, C\right\}_{\rm NB} + \left\{A, \frac{dB}{dt}, C\right\}_{\rm NB}
 + \left\{A, B, \frac{dC}{dt}\right\}_{\rm NB}
\label{D-rule-NB}
\end{eqnarray}
and the fundamental identity~\cite{Takhtajan}:
\begin{eqnarray}
\hspace{-1.2cm}
&~& \{\{A, G_1, G_2\}_{\rm NB}, H_1, H_2\}_{\rm NB}
\nonumber \\
\hspace{-1.2cm}
&~& ~~~= \{\{A, H_1, H_2\}_{\rm NB}, G_1, G_2\}_{\rm NB}
+ \{A, \{G_1, H_1, H_2\}_{\rm NB}, G_2\}_{\rm NB}
+ \{A, G_1, \{G_2, H_1, H_2\}_{\rm NB}\}_{\rm NB},
\label{F-id}
\end{eqnarray}
it is shown that the equation is invariant under 
the generalized canonical transformation such that
\begin{eqnarray}
\frac{dA'}{dt}=\{A', H_1, H_2\}_{\rm NB},
\label{dA'/dt-N}
\end{eqnarray}
where $A' = A + \delta_{G}A$.
From eqs.~\eqref{dA/dt-N} and \eqref{deltaG-N}, we see that the time evolution
of variables is realized by the generalized canonical transformation
whose generation functions are the Hamiltonians.

Here, we study the relationship between Nambu dynamics and Hamiltonian dynamics.
Since $H_1$ and $H_2$ conserve, a particle moves on the path
as the intersection of two surfaces described by $H_1=E_1$ and $H_2=E_2$.
If we choose $z=H_2$,
the Nambu equation becomes as
\begin{eqnarray}
\frac{dx}{dt}=\frac{\partial H_1}{\partial y},~~
\frac{dy}{dt}=-\frac{\partial H_1}{\partial x},~~
\frac{dz}{dt}=0,
\label{N-eq-H-eq}
\end{eqnarray}
and it turns out to be Hamilton's canonical equations of motion
by identifying $x$, $y$ and $H_1$ as $q$, $p$ and $H$, respectively.

Let us consider a specific system with variables $\theta$, $J_1$ and $J_2$,
in which they follow the Nambu equation and $\theta$ becomes a cyclic coordinate.
In this system, as Hamiltonians $H'_1$ and $H'_2$ do not depend on $\theta$,
the triplet satisfies the equation:
\begin{eqnarray}
\frac{d\theta}{dt}=\frac{\partial (H'_1, H'_2)}{\partial (J_1, J_2)},~~
\frac{dJ_1}{dt}=\frac{\partial (H'_1, H'_2)}{\partial (J_2, \theta)} = 0,~~
\frac{dJ_2}{dt}=\frac{\partial (H'_1, H'_2)}{\partial (\theta, J_1)} = 0.
\label{dtheta/dt-N}
\end{eqnarray}
From eq.~\eqref{dtheta/dt-N},
we find that $J_1$ and $J_2$ become constants and then $\theta=\theta(t)$ is given by
\begin{eqnarray}
\theta(t) = \nu t +\theta_0,
\label{theta(t)-N}
\end{eqnarray}
where $\theta_0$ is a constant and $\nu$ is also a constant given by
\begin{eqnarray}
\nu = \frac{\partial(E_1, E_2)}{\partial(J_1, J_2)}.
\label{nu-N}
\end{eqnarray} 
In eq.~\eqref{nu-N}, $E_1=E_1(J_1, J_2)$ and $E_2=E_2(J_1, J_2)$ 
are the values of $H'_1$ and $H'_2$, respectively.

If two systems described by eqs.~\eqref{N-eq} and \eqref{dtheta/dt-N}
are related to the generalized canonical transformation $(x, y, z) \to (\theta, J_1, J_2)$,
$\partial(\theta, J_1, J_2)/\partial(x, y, z) =1$ holds and
the canonical triplets are related to each other in the following way,
\begin{eqnarray}
x dy \wedge dz -  H_1 dH_2 \wedge dt 
= \theta dJ_1 \wedge dJ_2 - H'_1 dH'_2 \wedge dt + dW_{(1)},
\label{thetaJ1J2}
\end{eqnarray}
where $W_{(1)}$ is a 1-form.
The period $T$ is given by the relation:
\begin{eqnarray}
T = \int_0^T dt = \oint_{\rm C} \frac{1}{{dx}/{dt}} dx 
= \oint_{\rm C} \frac{1}{{\partial (H_1, H_2)}/{\partial (y, z)}} dx 
= \oint_{\rm C} \frac{\partial (y, z)}{\partial (E_1, E_2)} dx,
\label{T-N}
\end{eqnarray}
where C is a closed loop on the three-dimensional phase space, and
we use the relation $({\partial (y, z)}/{\partial (E_1, E_2)})
({\partial (H_1, H_2)}/{\partial (y, z)})=1$
and $dH_1 \wedge dH_2 = dH'_1 \wedge dH'_2
= dE_1 \wedge dE_2$.
When $\nu$ is regarded as a frequency, the following relation is obtained
\begin{eqnarray}
T = \frac{1}{\nu} = \frac{\partial(J_1, J_2)}{\partial(E_1, E_2)}
= \oint_{\rm C} \frac{\partial (y, z)}{\partial (E_1, E_2)} dx,
\label{1/nu-N}
\end{eqnarray}
using eqs.~\eqref{nu-N} and \eqref{T-N}.
Integrating eq.~\eqref{1/nu-N} over $E_1$ and $E_2$, we derive the relation:
\begin{eqnarray}
\iint \frac{dE_1 \wedge dE_2}{\nu} = \iint dJ_1 \wedge dJ_2
= \iiint dy \wedge dz \wedge dx.
\label{E1E2/nu-N}
\end{eqnarray}
It is natural to impose the following quantization condition 
on the canonical triplet,
\begin{eqnarray}
\iiint dx \wedge dy \wedge dz = hn,
\label{BSlike}
\end{eqnarray}
as a reference of the condition \eqref{BS} in the old quantum theory.

We have employed a generalized version of the Hamilton-Jacobi method for a system 
involving a cyclic coordinate; however, the applicability of our method is limited 
to specific classes of systems. 
Here, as an example, we consider the system described by the Hamiltonians:
\begin{eqnarray}
H_1 = (a_1 x + b_1 y)^2 + c_1 z^2 + f(a_2 x + b_2 y + z),~~ H_2 = a_2 x + b_2 y + z,
\label{Hs-ex}
\end{eqnarray}
where $a_1$, $b_1$, $c_1$, $a_2$ and $b_2$ are constants with $a_1 b_2 - a_2 b_1 >0$
and $c_1 > 0$.
After the generalized canonical transformation such as
\begin{eqnarray}
x' = \frac{1}{\sqrt{a_1 b_2 - a_2 b_1}}(a_1 x + b_1 y),~~
y' = \frac{1}{\sqrt{a_1 b_2 - a_2 b_1}}(a_2 x + b_2 y),~~
z' = a_2 x + b_2 y + z,
\label{x'-ex}
\end{eqnarray}
the above Hamiltonians are written by
\begin{eqnarray}
H_1 = (a_1 b_2 - a_2 b_1) x'^2 + c_1(a_1 b_2 - a_2 b_1)\left(\frac{z'}{\sqrt{a_1 b_2 - a_2 b_1}}
-y'\right)^2 + f(z'),~~ H_2 = z',
\label{H's-ex}
\end{eqnarray}
and this system is equivalent to one-dimensional harmonic oscillator system.
In fact, $J_1$ and $J_2$ are expressed as
\begin{eqnarray}
J_1 = \frac{\pi (E_1 - f(J_2))}{\sqrt{c_1}(a_1 b_2 - a_2 b_1)},~~
J_2 = E_2
\label{Js-ex}
\end{eqnarray}
and $\nu$ is given by
\begin{eqnarray}
\nu = \frac{\partial(E_1, E_2)}{\partial(J_1, J_2)} = \frac{\partial E_1}{\partial J_1} 
= \frac{\sqrt{c_1}(a_1b_2-a_2b_1)}{\pi}
\label{nu-N-ex}
\end{eqnarray} 
where $E_1$ and $E_2$ are the values of $H_1$ and $H_2$, respectively.
Incidentally, a generalization of the Hamilton-Jacobi theory that can be applied to 
any Nambu mechanical system has been formulated elegantly in Ref.~\cite{Yoneya}
using the mathematical concept of fiber bundles.

\subsection{Generalized Heisenberg dynamics}

Nambu examined how to quantize Nambu mechanics
and found it to be difficult~\cite{Nambu}.
After that various attempts on the quantization of Nambu mechanics
or the Nambu bracket have been carried out~\cite{Takhtajan,DFS&T,ALM&Y,Xiong,C&Z,H&M2}.

Let us formulate an extension of Heisenberg dynamics
from a specific class of Nambu dynamics
described by eq.~\eqref{dtheta/dt-N}, using the correspondence principle (see Figure 2).

\begin{figure}[htbp]
\includegraphics[width=10cm,bb=0 0 600 540]{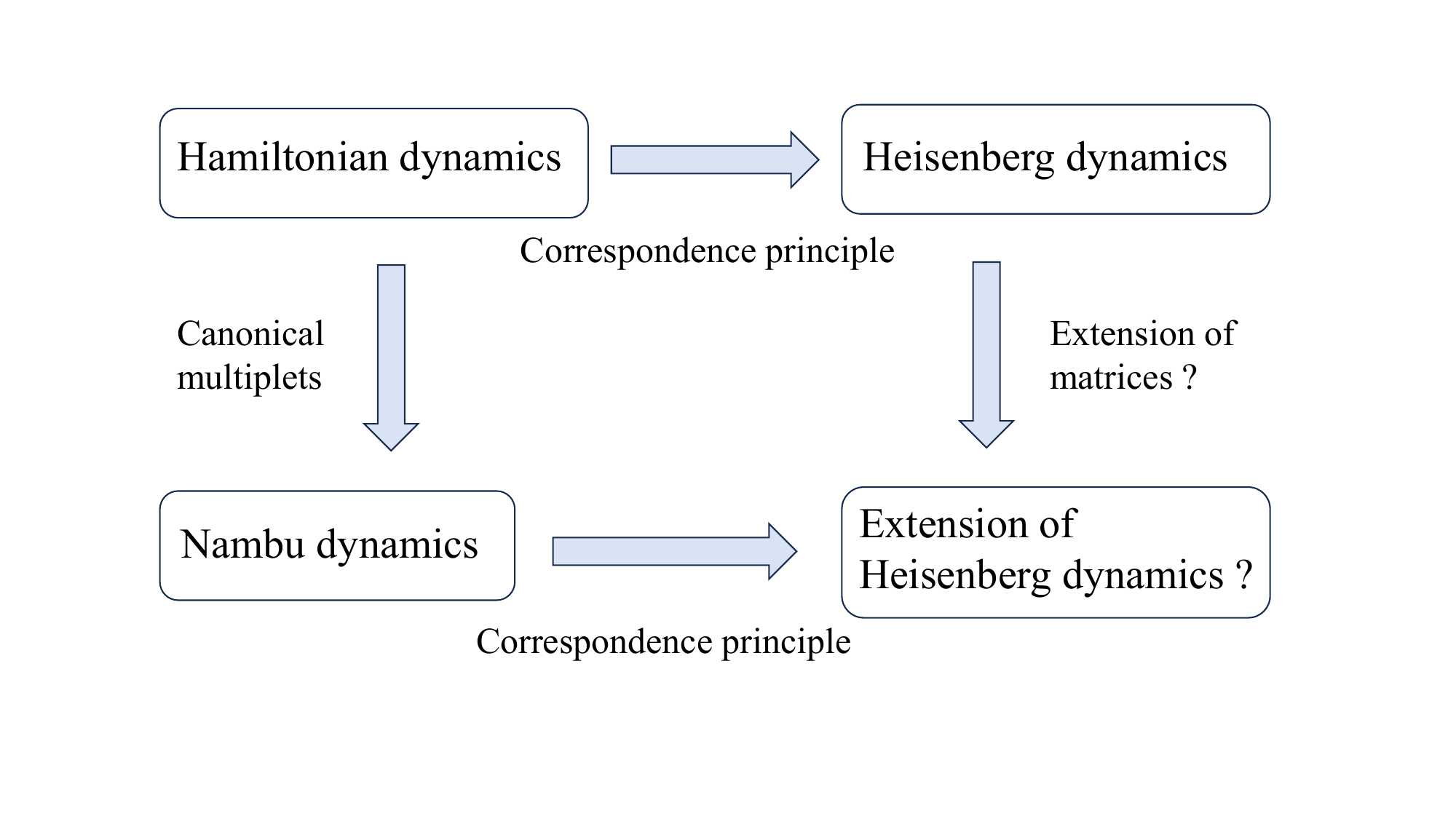}
\caption[F2]{Strategy for exploring new dynamics.}
\label{F2}
\end{figure}

With the system described by eq.~\eqref{dtheta/dt-N} in mind,
we assume that there is a one-to-one correspondence 
between values of $E_1(J_1, J_2)$ and $E_2(J_1, J_2)$ in Nambu dynamics
and some discrete values $(E_1)_{l~\!\! m}$ and $(E_2)_{l~\!\! m}$
in its extension,
and these discrete values are anti-symmetric such as
\begin{eqnarray}
(E_1)_{m~\!\! l} = -(E_1)_{l~\!\! m},~~ (E_2)_{m~\!\! l} = -(E_2)_{l~\!\! m}
\label{anti-sym}
\end{eqnarray}
and follow the combination rule:
\begin{eqnarray}
&~& (E_1)_{l~\!\! l-\Delta l}=(E_1)_{l~\!\! m} + (E_1)_{m~\!\! l-\Delta l}
= (E_1)_{l~\!\! m} - (E_1)_{l-\Delta l~\!\! m},~~
\label{R-E1}\\
&~& (E_2)_{m~\!\! m-\Delta m}=(E_2)_{m~\!\! l} + (E_2)_{l~\!\! m-\Delta m}
= (E_2)_{m~\!\! l} - (E_2)_{m-\Delta m~\!\! l},
\label{R-E2}
\end{eqnarray}
where $\Delta l$ and $\Delta m$ are natural numbers.
In this case, we conjecture the correspondence relation:
\begin{eqnarray}
&~& \left(\frac{(E_1)_{l~\!\! l-\Delta l}}{\Delta l}\frac{(E_2)_{m~\!\! m-\Delta m}}{\Delta m}
- \frac{(E_2)_{l~\!\! l-\Delta l}}{\Delta l}\frac{(E_1)_{m~\!\! m-\Delta m}}{\Delta m}\right)
\Delta l \Delta m
\nonumber\\
&~& ~~~~~
= \left(\frac{(E_1)_{l~\!\! m} - (E_1)_{l-\Delta l~\!\! m}}{\Delta l}
\frac{(E_2)_{m~\!\! l} - (E_2)_{m-\Delta m~\!\! l}}{\Delta m}\right.
\nonumber \\
&~& ~~~~~~~~~~~ \left.- \frac{(E_2)_{l~\!\! m} - (E_2)_{l-\Delta l~\!\! m}}{\Delta l}
\frac{(E_1)_{m~\!\! l} - (E_1)_{m-\Delta m~\!\! l}}{\Delta m}\right)
\Delta l \Delta m
\nonumber\\
&~& ~~~~~ \xrightarrow{l \gg \Delta l,~m \gg \Delta m} 
\frac{\partial(E_1, E_2)}{\partial(J_1, J_2)} \Delta l \Delta m 
= \nu \Delta l \Delta m,
\label{nu-N-CoPri}
\end{eqnarray}
and obtain a frequency condition such as
\begin{eqnarray}
\nu^{(0)}_{l~\!\! m~\!\! n} 
= \beta\left\{(E_1)_{l~\!\! n}(E_2)_{m~\!\! n}-(E_2)_{l~\!\! n}(E_1)_{m~\!\! n}\right\},
\label{nu-N-qu-0}
\end{eqnarray}
where $\beta$ is a constant and we choose $n=l-\Delta l=m-\Delta m$.
Eq.~\eqref{nu-N-qu-0} is regarded as a counterpart of the relation \eqref{nu-N}.

Using eqs.~\eqref{anti-sym}, \eqref{R-E1}, \eqref{R-E2} and \eqref{nu-N-qu-0}, 
we derive the relation:
\begin{eqnarray}
\nu^{(0)}_{l~\!\! m~\!\! n} = \nu^{(0)}_{l~\!\! m~\!\! k}
+ \nu^{(0)}_{m~\!\! n~\!\! k} + \nu^{(0)}_{n~\!\! l~\!\! k},
\label{nu-N-qu-rel}
\end{eqnarray}
which represents that $\nu^{(0)}_{l~\!\! m~\!\! n}$ is a 3-cocycle, i.e., 
$(\delta\nu^{(0)})_{l~\!\! m~\!\! n~\!\! k}=0$.
Here, $\delta$ is a coboundary operator defined by
\begin{eqnarray}
(\delta\nu^{(0)})_{l~\!\! m~\!\! n~\!\! k}
\equiv \nu^{(0)}_{m~\!\! n~\!\! k} - \nu^{(0)}_{l~\!\! n~\!\! k}
+ \nu^{(0)}_{l~\!\! m~\!\! k} - \nu^{(0)}_{l~\!\! m~\!\! n}.
\label{delta}
\end{eqnarray}
From eqs.~\eqref{nu-N-qu-0} and \eqref{nu-N-qu-rel}, 
we find that $\nu^{(0)}_{l~\!\! m~\!\! n}$ is totally anti-symmetric, i.e.,
$\nu^{(0)}_{l~\!\! m~\!\! n}=\nu^{(0)}_{m~\!\! n~\!\! l}=\nu^{(0)}_{n~\!\! l~\!\! m}
=-\nu^{(0)}_{m~\!\! l~\!\! n}=-\nu^{(0)}_{l~\!\! n~\!\! m}=-\nu^{(0)}_{n~\!\! m~\!\! l}$.

For later convenience, we adopt a manifestly cyclic form such as
\begin{align}
\nu_{l~\!\! m~\!\! n} 
&= \beta \left\{(E_1)_{l~\!\! n}(E_2)_{m~\!\! n}-(E_2)_{l~\!\! n}(E_1)_{m~\!\! n}
\right. + (E_1)_{m~\!\! l}(E_2)_{n~\!\! l}-(E_2)_{m~\!\! l}(E_1)_{n~\!\! l}
\nonumber \\
&~~~~ \left. 
+ (E_1)_{n~\!\! m}(E_2)_{l~\!\! m}-(E_2)_{n~\!\! m}(E_1)_{l~\!\! m}\right\},
\label{nu-N-qu}
\end{align}
as a frequency condition.
We note that $\nu_{l~\!\! m~\!\! n}$ is also totally anti-symmetric
and satisfies a generalization of Ritz's combination rule:
\begin{eqnarray}
\nu_{l~\!\! m~\!\! n} = \nu_{l~\!\! m~\!\! k} 
+ \nu_{l~\!\! k~\!\! n} + \nu_{k~\!\! m~\!\! n}.
\label{gR-rule}
\end{eqnarray}

Next, let us derive an equation of motion for variables with three indices
based on the condition \eqref{nu-N-qu}.
We consider a periodic motion with the period $T$.
Since $\nu_{l~\!\! m~\!\! n}$ has three indices,
we suppose that a dynamical variable containing it as a frequency also owns three indices.
Then, the quantities $(A_1)_{l~\!\! m~\!\! n}(t)$, $(A_2)_{l~\!\! m~\!\! n}(t)$ 
and $(A_3)_{l~\!\! m~\!\! n}(t)$ have a feature
that $(A_1)_{l~\!\! m~\!\! n}(t+T)=(A_1)_{l~\!\! m~\!\! n}(t)$, 
$(A_2)_{l~\!\! m~\!\! n}(t+T)=(A_2)_{l~\!\! m~\!\! n}(t)$ ,
and $(A_3)_{l~\!\! m~\!\! n}(t+T)=(A_3)_{l~\!\! m~\!\! n}(t)$
and can be described as
\begin{eqnarray}
&~& (A_1)_{l~\!\! m~\!\! n}(t)=(A_1)_{l~\!\! m~\!\! n}(0) e^{2\pi i \nu_{l~\!\! m~\!\! n}t},~~
(A_2)_{l~\!\! m~\!\! n}(t)=(A_2)_{l~\!\! m~\!\! n}(0) e^{2\pi i \nu_{l~\!\! m~\!\! n}t},~~
\nonumber \\
&~& (A_3)_{l~\!\! m~\!\! n}(t)=(A_3)_{l~\!\! m~\!\! n}(0) e^{2\pi i \nu_{l~\!\! m~\!\! n}t},
\label{Almn}
\end{eqnarray}
where $\nu_{l~\!\! m~\!\! n}$ follows the condition \eqref{nu-N-qu}
and is also expressed by $\nu_{l~\!\! m~\!\! n} = n_{l~\!\! m~\!\! n}/T$ 
($n_{l~\!\! m~\!\! n}$ : an integer). 

If we require that the product $(A_1 A_2 A_3)_{l~\!\! m~\!\! n}(t)$ 
should take the same form as eq.~\eqref{Almn}, i.e.,
$(A_1 A_2 A_3)_{l~\!\! m~\!\! n}(t)
=(A_1 A_2 A_3)_{l~\!\! m~\!\! n}(0) e^{2\pi i \nu_{l~\!\! m~\!\! n}t}$,
$(A_1)_{l~\!\! m~\!\! n}(t)$, $(A_2)_{l~\!\! m~\!\! n}(t)$ and $(A_3)_{l~\!\! m~\!\! n}(t)$ 
become components of extended matrices whose product is defined by
\begin{align}
(A_1 A_2 A_3)_{l~\!\! m~\!\! n}(t) 
\equiv \sum_k (A_1)_{l~\!\! m~\!\! k}(t) (A_2)_{l~\!\! k~\!\! n}(t) (A_3)_{k~\!\! m~\!\! n}(t),
\label{ABC}
\end{align}
where we use the generalized rule \eqref{gR-rule}.
We refer to three-index variables, which follow the product rule~\eqref{ABC},
as cubic matrices\footnote{
Three-index objects have been introduced to construct a quantum version of
the Nambu bracket in Ref.~\cite{ALM&Y}.
Their definition of the 3-fold product is different from ours.}.

Because the components with $l = m$, $m=n$ and/or $n=l$ on 
$(A_1)_{l~\!\! m~\!\! n}(t)$, $(A_2)_{l~\!\! m~\!\! n}(t)$ and $(A_3)_{l~\!\! m~\!\! n}(t)$
do not depend on time
as seen from eqs.~\eqref{nu-N-qu} and \eqref{Almn}, any conserved quantity 
satisfying $dC_{l~\!\! m~\!\! n}/dt = 0$ can be written as
\begin{eqnarray}
C_{l~\!\! m~\!\! n} = c_{l~\!\! m}(1-\delta_{l~\!\! n})\delta_{m~\!\! n} 
+ c_{m~\!\! n}(1-\delta_{m~\!\! l})\delta_{n~\!\! l} 
+ c_{n~\!\! l}(1-\delta_{n~\!\! m})\delta_{l~\!\! m},
\label{A(N)}
\end{eqnarray}
where $c_{l~\!\! m}$, $c_{m~\!\! n}$ and $c_{n~\!\! l}$ are constants.
We refer to the form of cubic matrices \eqref{A(N)} as a normal form
or a normal cubic matrix\footnote{Although a normal matrix is defined by 
$C_{l~\!\! m~\!\! n} = \delta_{l~\!\! m}c_{m~\!\! n} 
+ \delta_{m~\!\! n}c_{n~\!\! l} 
+ \delta_{n~\!\! l}c_{l~\!\! m}$ in Refs.~\cite{YK2,YK3},
we change the definition 
in order to match a definition of the normal form on generalized matrices 
with $n (\ge 3)$ indices presented in subsection 4.1.
In eq.~\eqref{A(N)}, we add extra factors such as $(1-\delta_{l~\!\! n})$,
$(1-\delta_{m~\!\! l})$ and $(1-\delta_{n~\!\! m})$
to meet the condition $C_{l~\!\! l~\!\! l} = 0$.
We note that the condition $C_{l~\!\! l~\!\! l} = 0$
is fulfilled without extra factors if $c_{l~\!\! m}$, $c_{m~\!\! n}$
and $c_{n~\!\! l}$ are anti-symmetric.}.

In a closed system, the Hamiltonians conserve, 
and hence $(H_1)_{l~\!\! m~\!\! n}$ 
and $(H_2)_{l~\!\! m~\!\! n}$ take the normal form:
\begin{eqnarray}
&~& (H_1)_{l~\!\! m~\!\! n} = 
(h_1)_{l~\!\! m}(1-\delta_{l~\!\! n})\delta_{m~\!\! n} 
+ (h_1)_{m~\!\! n}(1-\delta_{m~\!\! l})\delta_{n~\!\! l}
+ (h_1)_{n~\!\! l}(1-\delta_{n~\!\! m})\delta_{l~\!\! m},
\label{H1(N)}\\
&~& (H_2)_{l~\!\! m~\!\! n} = 
(h_2)_{l~\!\! m}(1-\delta_{l~\!\! n})\delta_{m~\!\! n} 
+ (h_2)_{m~\!\! n}(1-\delta_{m~\!\! l})\delta_{n~\!\! l}
+ (h_2)_{n~\!\! l}(1-\delta_{n~\!\! m})\delta_{l~\!\! m},
\label{H2(N)}
\end{eqnarray}
where $(h_1)_{l~\!\! m}$, $(h_1)_{m~\!\! n}$, $(h_1)_{n~\!\! l}$,
$(h_2)_{l~\!\! m}$, $(h_2)_{m~\!\! n}$ and $(h_2)_{n~\!\! l}$ are constants.
Using eqs.~\eqref{nu-N-qu}, \eqref{gR-rule}, \eqref{Almn}, \eqref{ABC},
\eqref{H1(N)} and \eqref{H2(N)}, 
we derive the equation:
\begin{align}
\frac{d}{dt} A_{l~\!\! m~\!\! n}(t) 
&= 2\pi i \nu_{l~\!\! m~\!\! n}A_{l~\!\! m~\!\! n}(t) 
\nonumber \\
&= 2\pi i \beta
\left\{(E_1)_{l~\!\! n}(E_2)_{m~\!\! n}-(E_2)_{l~\!\! n}(E_1)_{m~\!\! n}
+ (E_1)_{m~\!\! l}(E_2)_{n~\!\! l}-(E_2)_{m~\!\! l}(E_1)_{n~\!\! l}
\right.
\nonumber \\
&~~~~
\left. + (E_1)_{n~\!\! m}(E_2)_{l~\!\! m}-(E_2)_{n~\!\! m}(E_1)_{l~\!\! m}\right\}
A_{l~\!\! m~\!\! n}(t)
\nonumber \\
&= \frac{1}{i\hbar} \left[A(t), H_1, H_2\right]_{l~\!\! m~\!\! n},
\label{gH-eq-derive}
\end{align}
where we choose $\beta = -1/h$, $(E_1)_{l~\! n}=(h_1)_{l~\! n}$,
$(E_2)_{m~\! n}=(h_2)_{m~\! n}$ and so on. 
In eq.~\eqref{gH-eq-derive},
$\left[A(t), H_1, H_2\right]_{l~\!\! m~\!\! n}$ is a triple commutator defined by
\begin{align}
\left[A(t), H_1, H_2\right]_{l~\!\! m~\!\! n}
&\equiv \sum_{k} \left\{A(t)_{l~\!\! m~\!\! k} (H_1)_{l~\!\! k~\!\! n} (H_2)_{k~\!\! m~\!\! n}
+ (H_1)_{l~\!\! m~\!\! k} (H_2)_{l~\!\! k~\!\! n} A(t)_{k~\!\! m~\!\! n}\right.
\nonumber \\
&~~~~
+ (H_2)_{l~\!\! m~\!\! k} A(t)_{l~\!\! k~\!\! n} (H_1)_{k~\!\! m~\!\! n}
- (H_2)_{l~\!\! m~\!\! k} (H_1)_{l~\!\! k~\!\! n} A(t)_{k~\!\! m~\!\! n}
\nonumber \\
&~~~~
\left. 
- (H_1)_{l~\!\! m~\!\! k} A(t)_{l~\!\! k~\!\! n} (H_2)_{k~\!\! m~\!\! n}
- A(t)_{l~\!\! m~\!\! k} (H_2)_{l~\!\! k~\!\! n} (H_1)_{k~\!\! m~\!\! n}
\right\}.
\label{[A,H1,H2]}
\end{align}
In the derivation of eq.~\eqref{gH-eq-derive}, we use the relations
$(h_1)_{l~\! m}(1-\delta_{l~\! m}) = (h_1)_{l~\! m}$ 
and $(h_2)_{l~\! m}(1-\delta_{l~\! m}) = (h_2)_{l~\! m}$
derived from the anti-symmetric property
$(h_1)_{m~\! l}=-(h_1)_{l~\! m}$ and $(h_2)_{m~\! l}=-(h_2)_{l~\! m}$ (see eq.~\eqref{anti-sym}).

In this way, we have arrived at an extension of Heisenberg's equation of motion:
\begin{eqnarray}
\frac{d}{dt} A_{l~\!\! m~\!\! n}(t) 
= \frac{1}{i \hbar} \left[A(t), H_1, H_2\right]_{l~\!\! m~\!\! n},
\label{gH-eq}
\end{eqnarray}
as a counterpart of the Nambu equation \eqref{dA/dt-N}.
It is reconfirmed that the components with 
$l = m$, $m=n$ and/or $n=l$ on $A_{l~\!\! m~\!\! n}(t)$
are independent of time, using eq.~\eqref{gH-eq} and the relation
$[B, C_1, C_2]_{l~\!\! m~\!\! n}=0$ that holds concerning a cubic matrix $B$
with $B_{l~\!\! m~\!\! n}=0$ for $l \ne m$, $m \ne n$ and $n \ne l$
and arbitrary normal cubic matrices $C_1$ and $C_2$.
The equation \eqref{gH-eq} coincides with the fundamental equation of 
the generalized cubic matrix mechanics obtained in Ref.~\cite{YK3}
through its structural analogy with Nambu mechanics.

It is natural to impose the triplet of cubic matrices $X_{l~\!\! m~\!\! n}$,
$Y_{l~\!\! m~\!\! n}$ and $Z_{l~\!\! m~\!\! n}$ on the commutation relations:
$\left[X, Y, Z\right]_{l~\!\! m~\!\! n}= i\hbar I_{l~\!\! m~\!\! n}$
and others$=0$ (or anti-commutation relations for cubic matrices describing fermionic variables)
where $I$ is a specific constant cubic matrix (given by eq.~\eqref{I-def}).
In our generalized Heisenberg dynamics, the objective is to determine
the values of Hamiltonians, i.e., $(h_1)_{l~\!\! m}$, $(h_1)_{m~\!\! n}$, $(h_1)_{n~\!\! l}$,
$(h_2)_{l~\!\! m}$, $(h_2)_{m~\!\! n}$ and $(h_2)_{n~\!\! l}$ in eqs.~\eqref{H1(N)}
and \eqref{H2(N)},taking the equation of motion \eqref{gH-eq}
as the master equation.
However, since there is currently no method analogous to a unitary transformation
in matrix mechanics, 
these values must be obtained directly. 

Concerning the relationship between symmetries and conservation laws, 
cubic matrices $C^a$ that commute with the Hamiltonians, i.e.,
$[C^a, H_1, H_2] = 0$, are identified as conserved quantities, 
as understood from the generalized Heisenberg equation \eqref{gH-eq}.
Furthermore, arbitrary constant normal matrices become conserved quantities,
because the relation $[N_1, N_2, N_3]=0$ holds
for any normal cubic matrices $N_1$, $N_2$ and $N_3$.

Under the infinitesimal transformation defined by
\begin{eqnarray}
\delta_{G}A_{l~\!\! m~\!\! n}=\varepsilon [A, G_1, G_2]_{l~\!\! m~\!\! n}
= \varepsilon \widetilde{(G_1G_2)}_{l~\!\! m~\!\! n}A_{l~\!\! m~\!\! n}
\label{deltaG-gH}
\end{eqnarray}
with an infinitesimal constant $\varepsilon$
and normal cubic matrices $(G_1)_{l~\!\! m~\!\! n}$ and $(G_2)_{l~\!\! m~\!\! n}$,
using the derivation rule:
\begin{eqnarray}
\frac{d}{dt} [A, B, C]_{l~\!\! m~\!\! n} = 
\left[\frac{dA}{dt}, B, C\right]_{l~\!\! m~\!\! n} 
+ \left[A, \frac{dB}{dt}, C\right]_{l~\!\! m~\!\! n}
+ \left[A, B, \frac{dC}{dt}\right]_{l~\!\! m~\!\! n}
\label{D-rule-gH}
\end{eqnarray}
and a cubic matrix version of the fundamental identity\footnote{
When normal forms play the role of generators,
the validity of the fundamental identity is confirmed for generalized matrices
with $n (\ge 3)$ indices in subsection 4.2.
In eq.~\eqref{F-id-gH}, the Hamiltonians $H_1$ and $H_2$ act as generators.}:
\begin{eqnarray}
\hspace{-1.2cm}
&~& [[A, G_1, G_2], H_1, H_2]_{l~\!\! m~\!\! n}
\nonumber \\
\hspace{-1.2cm}
&~& ~~~= [[A, H_1, H_2], G_1, G_2]_{l~\!\! m~\!\! n}
+ [A, [G_1, H_1, H_2], G_2]_{l~\!\! m~\!\! n}
+ [A, G_1, [G_2, H_1, H_2]]_{l~\!\! m~\!\! n},
\label{F-id-gH}
\end{eqnarray}
it is shown that the equation \eqref{gH-eq} is invariant such that
\begin{eqnarray}
\frac{d}{dt}A'_{l~\!\! m~\!\! n}=[A', H_1, H_2]_{l~\!\! m~\!\! n},
\label{dA'/dt-gH}
\end{eqnarray}
where $A'_{l~\!\! m~\!\! n} = A_{l~\!\! m~\!\! n} + \delta_{G}A_{l~\!\! m~\!\! n}$.
In eq.~\eqref{deltaG-gH}, $\widetilde{(G_1G_2)}_{l~\!\! m~\!\! n}$ is defined by
\begin{align}
\widetilde{(G_1G_2)}_{l~\!\! m~\!\! n}
&\equiv (g_1)_{l~\!\! n}(g_2)_{m~\!\! n}-(g_2)_{l~\!\! n}(g_1)_{m~\!\! n}
+ (g_1)_{m~\!\! l}(g_2)_{n~\!\! l}-(g_2)_{m~\!\! l}(g_1)_{n~\!\! l}
\nonumber \\
&~~~~ + (g_1)_{n~\!\! m}(g_2)_{l~\!\! m}-(g_2)_{n~\!\! m}(g_1)_{l~\!\! m},
\label{tildeGG}
\end{align}
using normal cubic matrices $(G_1)_{l~\!\! m~\!\! n}$ and $(G_2)_{l~\!\! m~\!\! n}$
given by
\begin{eqnarray}
&~& (G_1)_{l~\!\! m~\!\! n} = 
(g_1)_{l~\!\! m}(1-\delta_{l~\!\! n})\delta_{m~\!\! n} 
+ (g_1)_{m~\!\! n}(1-\delta_{m~\!\! l})\delta_{n~\!\! l}
+ (g_1)_{n~\!\! l}(1-\delta_{n~\!\! m})\delta_{l~\!\! m},
\label{G1(N)}\\
&~& (G_2)_{l~\!\! m~\!\! n} = 
(g_2)_{l~\!\! m}(1-\delta_{l~\!\! n})\delta_{m~\!\! n} 
+ (g_2)_{m~\!\! n}(1-\delta_{m~\!\! l})\delta_{n~\!\! l}
+ (g_2)_{n~\!\! l}(1-\delta_{n~\!\! m})\delta_{l~\!\! m},
\label{G2(N)}
\end{eqnarray}
where $(g_1)_{m~\! l}$s and $(g_2)_{m~\! l}$s are anti-symmetric constants.
Eq.~\eqref{dA'/dt-gH} is obtained in a similar way to the counterpart in the Nambu mechanics.
From eqs.~\eqref{gH-eq} and \eqref{deltaG-gH}, 
we find that the time evolution
of variables is generated by the transformation
whose generators are the Hamiltonians.

When we deal with eq.~\eqref{gH-eq} as the master equation,
we find that various systems can be described as solutions of this equation.
Here, we present another time-dependent solution satisfying 
the generalization of Ritz's combination rule \eqref{gR-rule}.
It is written by
\begin{eqnarray}
&~& A_{l~\!\! m~\!\! n}(t)=A_{l~\!\! m~\!\! n}(0) e^{2\pi i \tilde{\nu}_{l~\!\! m~\!\! n}t},
\label{Almn-an}\\
&~& \tilde{\nu}_{l~\!\! m~\!\! n} = \frac{2}{h}
\left\{(\tilde{h}_1)_{n~\!\! l}+(\tilde{h}_1)_{l~\!\! m}+(\tilde{h}_1)_{m~\!\! n}\right\},
\label{nulmn-an}
\end{eqnarray}
where $(\tilde{h}_1)_{l~\!\! m}$, $(\tilde{h}_1)_{m~\!\! n}$ and $(\tilde{h}_1)_{n~\!\! l}$
are anti-symmetric constants
but do not satisfy the combination rule such as eq.~\eqref{R-E1}, i.e.,
$(\tilde{h}_1)_{l~\!\! m} \ne (\tilde{h}_1)_{l~\!\! n}+(\tilde{h}_1)_{n~\!\! m}$.
As seen from eq.~\eqref{nulmn-an}, 
$\tilde{\nu}_{l~\!\! m~\!\! n}$ takes the form of 3-coboundary 
and satisfies the cocycle condition:
\begin{eqnarray}
(\delta\tilde{\nu})_{l~\!\! m~\!\! n~\!\! k}
= \tilde{\nu}_{m~\!\! n~\!\! k} - \tilde{\nu}_{l~\!\! n~\!\! k}
+ \tilde{\nu}_{l~\!\! m~\!\! k} - \tilde{\nu}_{l~\!\! m~\!\! n} = 0.
\label{cocycle-cond-an}
\end{eqnarray}
In this solution, $(H_1)_{l~\!\! m~\!\! n}$ 
and $(H_2)_{l~\!\! m~\!\! n}$ take the following normal forms:
\begin{eqnarray}
&~& (H_1)_{l~\!\! m~\!\! n} = 
(\tilde{h}_1)_{l~\!\! m}(1-\delta_{l~\!\! n})\delta_{m~\!\! n} 
+ (\tilde{h}_1)_{m~\!\! n}(1-\delta_{m~\!\! l})\delta_{n~\!\! l}
+ (\tilde{h}_1)_{n~\!\! l}(1-\delta_{n~\!\! m})\delta_{l~\!\! m},
\label{H1(N)-an}\\
&~& (H_2)_{l~\!\! m~\!\! n} = (1-\delta_{l~\!\! n})\delta_{m~\!\! n} 
+ (1-\delta_{m~\!\! l})\delta_{n~\!\! l} + (1-\delta_{n~\!\! m})\delta_{l~\!\! m}.
\label{H2(N)-an}
\end{eqnarray}
The present solution agrees with the solution of the fundamental equation of cubic matrix mechanics derived in Refs.~\cite{YK1,YK2}.
In the Appendix B, it is shown that the generalized matrix mechanics can describe
physical systems in conventional matrix mechanics.

\subsubsection{Fermionic harmonic oscillator}

We study a harmonic oscillator relating to fermionic variables~\cite{YK1} as an example
of a solution describing eqs.~\eqref{Almn-an} and \eqref{nulmn-an}. 
They are represented by the $3 \times 3 \times 3$ cubic matrices 
whose components are given by
\begin{eqnarray}
\xi_{l~\!\! m~\!\! n}(t) = \frac{1}{\sqrt{2}} |\varepsilon_{l~\!\! m~\!\! n}| 
e^{-i\omega \varepsilon_{l~\!\! m~\!\! n} t},~~
\eta_{l~\!\! m~\!\! n}(t) = -\frac{i}{\sqrt{2}} \varepsilon_{l~\!\! m~\!\! n}
e^{-i\omega \varepsilon_{l~\!\! m~\!\! n} t},
\label{xi-3}
\end{eqnarray}
where $l$, $m$ and $n$ run from $1$ to $3$, $\omega$ is the angular frequency,
$\varepsilon_{1~\!\! 2~\!\! 3}=\varepsilon_{2~\!\! 3~\!\! 1}=\varepsilon_{3~\!\! 1~\!\! 2}
=-\varepsilon_{3~\!\! 2~\!\! 1}=-\varepsilon_{2~\!\! 1~\!\! 3}=-\varepsilon_{1~\!\! 3~\!\! 2}
=1$ and others are zero. 
These variables satisfy the anti-commutation relations:
\begin{eqnarray}
\left\{\xi(t), I, \xi(t)\right\}_{l~\!\! m~\!\! n} = I_{l~\!\! m~\!\! n},~~
\left\{\eta(t), I, \eta(t)\right\}_{l~\!\! m~\!\! n} = I_{l~\!\! m~\!\! n},~~
\left\{\xi(t), I, \eta(t)\right\}_{l~\!\! m~\!\! n} = 0,
\label{antiCR-3}
\end{eqnarray}
where $\{A_1, A_2, A_3\}_{l~\!\! m~\!\! n}$ is the triple anti-commutator defined by
\begin{align}
\{A_1, A_2, A_3\}_{l~\!\! m~\!\! n}
&\equiv \sum_{k} \left\{(A_1)_{l~\!\! m~\!\! k} (A_2)_{l~\!\! k~\!\! n} (A_3)_{k~\!\! m~\!\! n}
+ (A_2)_{l~\!\! m~\!\! k} (A_3)_{l~\!\! k~\!\! n} (A_1)_{k~\!\! m~\!\! n}\right.
\nonumber \\
&~~~~
+ (A_3)_{l~\!\! m~\!\! k} (A_1)_{l~\!\! k~\!\! n} (A_2)_{k~\!\! m~\!\! n}
+ (A_2)_{l~\!\! m~\!\! k} (A_1)_{l~\!\! k~\!\! n} (A_3)_{k~\!\! m~\!\! n}
\nonumber \\
&~~~~
\left. 
+ (A_1)_{l~\!\! m~\!\! k} (A_3)_{l~\!\! k~\!\! n} (A_2)_{k~\!\! m~\!\! n}
+ (A_3)_{l~\!\! m~\!\! k} (A_2)_{l~\!\! k~\!\! n} (A_1)_{k~\!\! m~\!\! n}\right\},
\label{{A,H1,H2}}
\end{align}
and $I$ is a special type of normal form whose components are defined by
\begin{eqnarray}
I_{l~\!\! m~\!\! n} \equiv (1-\delta_{l~\!\! n})\delta_{m~\!\! n} 
+ (1-\delta_{m~\!\! l})\delta_{n~\!\! l} + (1-\delta_{n~\!\! m})\delta_{l~\!\! m}.
\label{I-def}
\end{eqnarray}
We find that $I$ can play the role of the identity
for the components $A_{l~\!\! m~\!\! n}$ $(l \ne m, m \ne n, n \ne l)$
from the relation 
$(AII)_{l~\!\! m~\!\! n} = (IAI)_{l~\!\! m~\!\! n} =(IIA)_{l~\!\! m~\!\! n}
= A_{l~\!\! m~\!\! n}$.

Fermionic variables 
become solutions of the generalized Heisenberg's equation of motion:
\begin{eqnarray}
\frac{d}{dt} \xi_{l~\!\! m~\!\! n}(t) = \frac{1}{i \hbar} [\xi(t), H_1, H_2]_{l~\!\! m~\!\! n},~~
\frac{d}{dt} \eta_{l~\!\! m~\!\! n}(t) = \frac{1}{i \hbar} [\eta(t), H_1, H_2]_{l~\!\! m~\!\! n},
\label{H-eq-xi-3}
\end{eqnarray}
where the Hamiltonians are given by
\begin{align}
(H_1)_{l~\!\! m~\!\! n} &= i\frac{\hbar\omega}{6} 
\left[\xi(t), I, \eta(t)\right]_{l~\!\! m~\!\! n}
= \frac{\hbar\omega}{6}
\left(\varepsilon_{l~\!\! m}\delta_{m~\!\! n}
+ \varepsilon_{m~\!\! n}\delta_{n~\!\! l}
+ \varepsilon_{n~\!\! l}\delta_{l~\!\! m}\right), 
\label{H1-xi}\\
(H_2)_{l~\!\! m~\!\! n} &= I_{l~\!\! m~\!\! n}
\label{H2-xi}
\end{align}
with the anti-symmetric quantity
$\varepsilon_{l~\!\! m} \equiv \sum_{k=1}^{3}\varepsilon_{l~\!\! m~\!\! k}$.
From eqs.~\eqref{Almn-an}, \eqref{xi-3} and \eqref{H-eq-xi-3}, 
$\tilde{\nu}_{l~\!\! m~\!\! n}$ is written as
\begin{eqnarray}
\tilde{\nu}_{l~\!\! m~\!\! n} = -\frac{\omega}{2\pi} \varepsilon_{l~\!\! m~\!\! n}
= -\frac{\omega}{6\pi}\left(\varepsilon_{n~\!\! l} + \varepsilon_{l~\!\! m}
+ \varepsilon_{m~\!\! n}\right),
\label{nu-3}
\end{eqnarray}
and it satisfies the cocycle condition $(\delta\tilde{\nu})_{l~\!\! m~\!\! n~\!\! k}=0$.
Since $\xi_{l~\!\! m~\!\! n}(t)$ and $\eta_{l~\!\! m~\!\! n}(t)$ follow the equations:
\begin{eqnarray}
\frac{d}{dt} \xi_{l~\!\! m~\!\! n}(t) = \omega {\eta}_{l~\!\! m~\!\! n}(t),~~
\frac{d}{dt} \eta_{l~\!\! m~\!\! n}(t) = - \omega {\xi}_{l~\!\! m~\!\! n}(t),
\label{dxi-3}
\end{eqnarray}
they obey the equation for simple harmonic motion:
\begin{eqnarray}
\frac{d^2}{dt^2} \xi_{l~\!\! m~\!\! n}(t) = - \omega^2 {\xi}_{l~\!\! m~\!\! n}(t),~~
\frac{d^2}{dt^2} \eta_{l~\!\! m~\!\! n}(t) = - \omega^2 {\eta}_{l~\!\! m~\!\! n}(t).
\label{d2xi-3}
\end{eqnarray}

The same system is described by the cubic matrices 
interpreted as the ladder operators:
\begin{eqnarray}
&~& C_{l~\!\! m~\!\! n}(t) \equiv \frac{1}{\sqrt{2}}\left(\xi_{l~\!\! m~\!\! n}(t) 
+ i \eta_{l~\!\! m~\!\! n}(t)\right)
= \frac{1}{2} \left(|\varepsilon_{l~\!\! m~\!\! n}| 
+ \varepsilon_{l~\!\! m~\!\! n}\right)e^{-i\omega t},~~
\label{C-3}\\
&~& C^{\dagger}_{l~\!\! m~\!\! n}(t) \equiv \frac{1}{\sqrt{2}}\left(\xi_{l~\!\! m~\!\! n}(t) 
- i \eta_{l~\!\! m~\!\! n}(t)\right)
= \frac{1}{2} \left(|\varepsilon_{l~\!\! m~\!\! n}| 
- \varepsilon_{l~\!\! m~\!\! n}\right)e^{i\omega t}.
\label{Cdag-3}
\end{eqnarray}
These variables satisfy the anti-commutation relations:
\begin{eqnarray}
\{C(t), I, C^{\dagger}(t)\}_{l~\!\! m~\!\! n} = I_{l~\!\! m~\!\! n},~~
\{C(t), I, C(t)\}_{l~\!\! m~\!\! n} = 0,~~
\{C^{\dagger}(t), I, C^{\dagger}(t)\}_{l~\!\! m~\!\! n} = 0,
\label{antiCR-C-3}
\end{eqnarray}
and become solutions of the generalized Heisenberg's equation of motion:
\begin{eqnarray}
\frac{d}{dt} C_{l~\!\! m~\!\! n}(t) = \frac{1}{i \hbar} [C(t), H_1, H_2]_{l~\!\! m~\!\! n},~~
\frac{d}{dt} C^{\dagger}_{l~\!\! m~\!\! n}(t) 
= \frac{1}{i \hbar} [C^{\dagger}(t), H_1, H_2]_{l~\!\! m~\!\! n},
\label{H-eq-C-3}
\end{eqnarray}
where the Hamiltonians are written by
\begin{eqnarray}
(H_1)_{l~\!\! m~\!\! n} 
= \frac{\hbar \omega}{6} [C^{\dagger}(t), I, C(t)]_{l~\!\! m~\!\! n},~~
(H_2)_{l~\!\! m~\!\! n} = I_{l~\!\! m~\!\! n}.
\label{H-C-3}
\end{eqnarray}
Since $C_{l~\!\! m~\!\! n}(t)$ and $C^{\dagger}_{l~\!\! m~\!\! n}(t)$ follow the equations:
\begin{eqnarray}
\frac{d}{dt} C_{l~\!\! m~\!\! n}(t) = -i\omega C_{l~\!\! m~\!\! n}(t),~~
\frac{d}{dt} C^{\dagger}_{l~\!\! m~\!\! n}(t) = i \omega C^{\dagger}_{l~\!\! m~\!\! n}(t),
\label{dC-3}
\end{eqnarray}
they also obey the equation for simple harmonic motion.

\section{Generalized Heisenberg dynamics of variables with $n$ indices}

\subsection{Generalized matrix mechanics}

Let us start from a generalization of eq.~\eqref{nu-N}:
\begin{eqnarray}
\nu = \frac{\partial(E_1, E_2, \cdots, E_{n-1})}{\partial(J_1, J_2, \cdots, J_{n-1})}
= \sum_{(a_1, a_2, \cdots, a_{n-1})}{\rm sgn(P)}
 \frac{\partial E_{a_1}}{\partial J_1} \frac{\partial E_{a_2}}{\partial J_2} \cdots
 \frac{\partial E_{a_{n-1}}}{\partial J_{n-1}},
\label{nu-N-n}
\end{eqnarray}
where sgn(P) is $+1$ and $-1$ for even and odd permutations among the set of subscripts
$(a_1, a_2, \cdots, a_{n-1})$, respectively, based on $+1$ 
for $(a_1, a_2, \cdots, a_{n-1})=(1, 2, \cdots, n-1)$,
and each $a_k$ $(k=1, 2, \cdots, n-1)$ runs from $1$ to $n-1$.
The above eq.~\eqref{nu-N-n} can be derived from Nambu dynamics describing
a specific periodic motion by the canonical $n$-plets 
($\theta$, $J_1$, $\cdots$, $J_{n-1}$) including a cyclic coordinate $\theta$
and the Hamiltonians ($H_1$, $H_2$, $\cdots$, $H_{n-1}$).
Then, as the Hamiltonians do not contain $\theta$,
the equations of motion are given by
\begin{eqnarray}
\hspace{-1cm}&~& \frac{d\theta}{dt}=
\frac{\partial (H_1, H_2, \cdots, H_{n-1})}{\partial (J_1, J_2, \cdots, J_{n-1})},~~
\label{dtheta-n}\\
\hspace{-1cm}&~& \frac{dJ_1}{dt}=
\frac{\partial (H_1, \cdots, H_{n-2}, H_{n-1})}{\partial (J_2, \cdots, J_{n-1}, \theta)} = 0,~
\cdots,~ 
\frac{dJ_{n-1}}{dt}=
\frac{\partial (H_1, H_2, \cdots, H_{n-1})}{\partial (\theta, J_1, \cdots, J_{n-2})}= 0.
\label{dJ-n}
\end{eqnarray}
From eqs.~\eqref{dtheta-n} and \eqref{dJ-n},
we find that $J_k$s become constants and then $\theta=\theta(t)$ is given by
$\theta(t) = \nu t +\theta_0$,
where $\theta_0$ is a constant and $\nu$ is also a constant given by eq.~\eqref{nu-N-n}
relating to $E_k$s (the values of $H_k$s), respectively.

In the following, 
we construct a mechanics of quantities with $n$ indices, 
using the correspondence principle,
or tackle a puzzle like a cryptarithm shown in Figure 3.

\begin{figure}[htbp]
\includegraphics[width=10cm,bb=0 0 600 540]{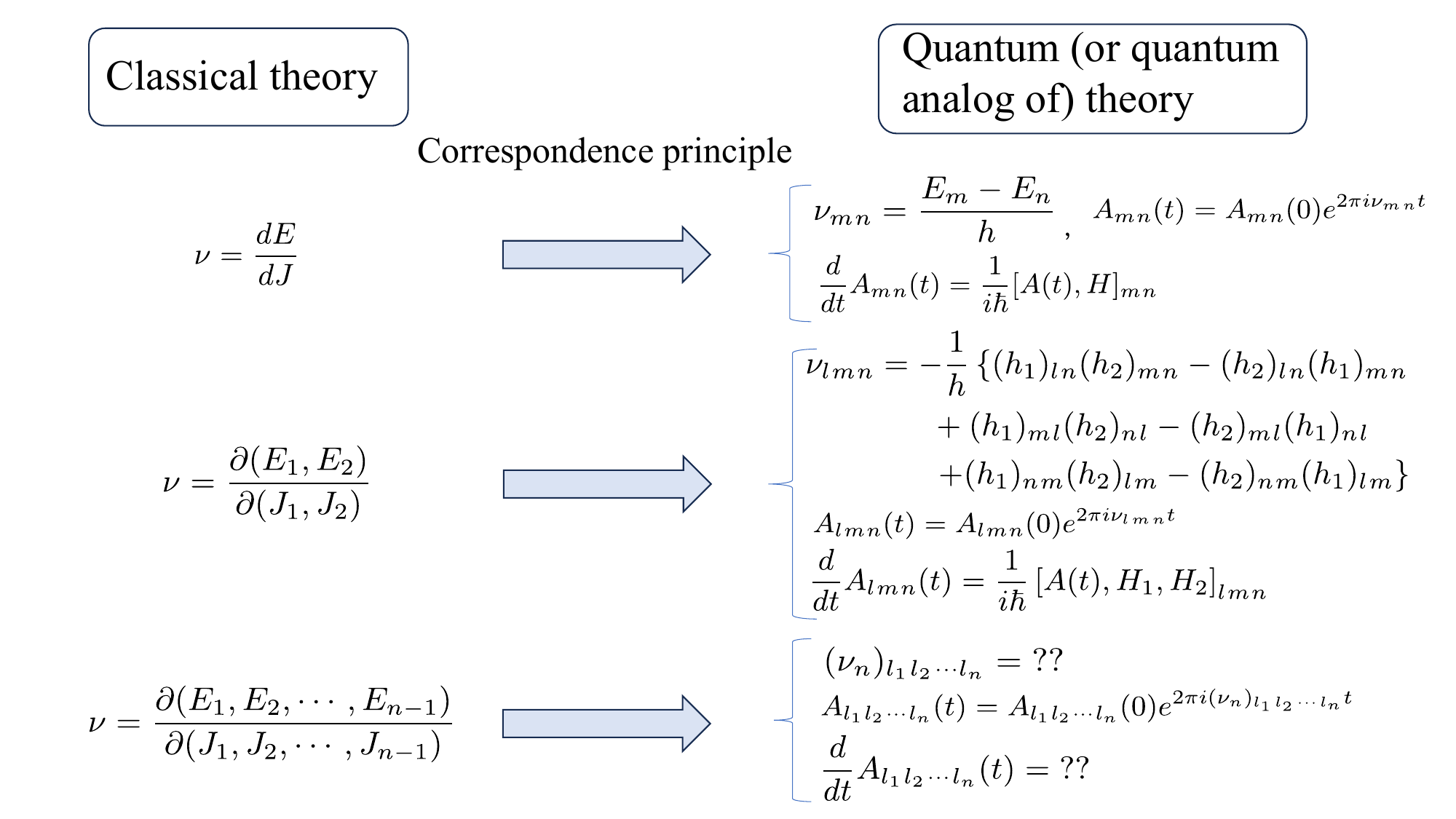}
\caption[F3]{Classical theory and its quantum (or quantum analog of) theory.}
\label{F3}
\end{figure}

We assume that there is a one-to-one correspondence 
between  $E_{a_k}$ in eq.~\eqref{nu-N-n}
and some discrete values $(E_{a_k})_{l_{r}~\!\! l_{r'}}$ 
($l_{r}$, $l_{r'}$ : natural numbers, $r, r' = 1, 2, \cdots, n$),
and these discrete values are anti-symmetric as
\begin{eqnarray}
(E_{a_k})_{l_{r'}~\!\! l_{r}} = -(E_{a_k})_{l_{r}~\!\! l_{r'}}
\label{anti-sym-n}
\end{eqnarray}
and follow the combination rule:
\begin{eqnarray}
(E_{a_k})_{l_{r}~\!\! l_{r'}}=(E_{a_k})_{l_{r}~\!\! l_{r''}}+(E_{a_k})_{l_{r''}~\!\! l_{r'}}
=(E_{a_k})_{l_{r}~\!\! l_{r''}}-(E_{a_k})_{l_{r'}~\!\! l_{r''}},
\label{R-E-n}
\end{eqnarray}
where $l_{r''}$ is also a natural number.

By identifying ${\partial E_{a_k}}/{\partial J_b}$ with 
${(E_{a_k})_{l_{b}~\!\! l_{n}}}/{\varDelta l_{b}}$ 
($l_n = l_b - \varDelta l_{b}$, $\varDelta l_{b}$ : a natural number, $b=1, 2, \cdots, n-1$),
we obtain the relation such as
\begin{eqnarray}
(\nu_n^{(0)})_{l_1~\!\! l_2~\!\! \cdots l_n} 
= \beta  \sum_{(a_1, a_2, \cdots, a_{n-1})}{\rm sgn(P)} 
\prod_{r=1}^{n-1}(E_{a_r})_{l_r~\!\! l_n},
\label{nu-N-qu-0-n}
\end{eqnarray}
where $\beta$ is a constant, and $(\nu_n^{(0)})_{l_1~\!\! l_2~\!\! \cdots l_n}$
is regarded as a counterpart of $\nu \varDelta l_1 \varDelta l_2 \cdots \varDelta l_{n-1}$.
From eq.~\eqref{nu-N-qu-0-n}, we see that 
$(\nu_n^{(0)})_{l_1~\!\! l_2~\!\! \cdots l_n}$ is anti-symmetric
under the exchange of $n-1$ indices $(l_1, l_2, \cdots, l_{n-1})$.

Using eqs.~\eqref{anti-sym-n}, \eqref{R-E-n} and \eqref{nu-N-qu-0-n}, 
we derive the relation\footnote{
The details of derivation of this relation and most other equations 
are presented in Appendix A.}:
\begin{eqnarray}
(\nu_n^{(0)})_{l_1~\!\! l_2~\!\! \cdots~\!\! l_n}
= \sum_{i = 1}^n (-1)^{n-i} (\nu_n^{(0)})_{l_1~\!\! \cdots~\!\! \hat{l}_{i}~\!\!
 \cdots~\!\! l_n~\!\! l_{n+1}},
\label{nu-N-qu-rel-n}
\end{eqnarray}
where the hatted index $\hat{l}_i$ means that $l_i$ is omitted,
and $l_{n+1}$ is a natural number.
Eq.~\eqref{nu-N-qu-rel-n} implies
that $(\nu_n^{(0)})_{l_1~\!\! l_2~\!\! \cdots~\!\! l_n}$ is a $n$-cocycle, i.e., 
$(\delta\nu_n^{(0)})_{l_1~\!\! l_2~\!\! \cdots~\!\! l_{n+1}}=0$.
Here, $\delta$ is a coboundary operator defined by
\begin{eqnarray}
(\delta\nu_n^{(0)})_{l_1~\!\! l_2~\!\! \cdots~\!\! l_{n+1}}
= \sum_{i=1}^{n+1} (-1)^{i+1} 
(\nu_n^{(0)})_{l_1~\!\! \cdots~\!\! \hat{l}_i~\!\! \cdots~\!\! l_{n+1}}.
\label{delta-n}
\end{eqnarray}

Let us require that $(\nu_n)_{l_1~\!\! l_2~\!\! \cdots~\!\! l_n}$
should have a cyclic property among indices\footnote{
The frequency $(\nu_n^{(0)})_{l_1~\!\! l_2~\!\! \cdots~\!\! l_n}$ 
given by eq.~\eqref{nu-N-qu-0-n} also have a cyclic property in a hidden way, 
as shown in Appendix A, and hence we find that it is totally
anti-symmetric under the exchange of $(l_1, l_2, \cdots, l_{n})$.}
such as
\begin{eqnarray}
(\nu_n)_{l_{p+1}~\!\! l_{p+2}~\!\! \cdots~\!\! l_{n}~\!\! l_1~\!\! \cdots~\!\! l_{p}}
= (-1)^{(n-1)p} (\nu_n)_{l_1~\!\! l_2~\!\! \cdots~\!\! l_n},
\label{nu-cyclic-n}
\end{eqnarray}
in a manifest way.
Here, we use the notation such as $l_{p} = l_{q}$ with $p=q$ (mod $n$)
and do it hereafter.
The following type of frequency possesses a manifestly cyclic symmetry
\begin{eqnarray}
(\nu_n)_{l_1~\!\! l_2~\!\! \cdots~\!\! l_n}
= \sum_{k=1}^{n} (-1)^{(n-1)k} 
(\nu_n^{(0)})_{l_{k+1}~\!\! l_{k+2}~\!\! \cdots~\!\! l_{n}~\!\! l_1~\!\! \cdots~\!\! l_{k}},
\label{nu-N-qu-n}
\end{eqnarray}
and then it is also totally anti-symmetric
under the exchange of $(l_1, l_2, \cdots, l_{n})$.

Using eqs.~\eqref{nu-N-qu-0-n} and \eqref{nu-N-qu-n}, we obtain the relation:
\begin{eqnarray}
(\nu_n)_{l_1~\!\! l_2~\!\! \cdots l_n} 
= \beta \sum_{k=1}^{n} (-1)^{n-k} \sum_{(a_1, a_2, \cdots, a_{n-1})}{\rm sgn(P)} 
\prod_{r=1}^{k-1}(E_{a_r})_{l_r~\!\! l_k}\prod_{r'=k}^{n-1}(E_{a_{r'}})_{l_{r'+1}~\!\! l_k}.
\label{nu-N-qu-n-E}
\end{eqnarray}
We find that eq.~\eqref{nu-N-qu-rel-n} holds for 
$(\nu_n)_{l_1~\!\! l_2~\!\! \cdots l_n}$ 
and $(\nu_n)_{l_1~\!\! l_2~\!\! \cdots l_n}$ also becomes $n$-cocycle
and satisfies a generalization of Ritz's combination rule:
\begin{eqnarray}
(\nu_n)_{l_1~\!\! l_2~\!\! \cdots~\!\! l_n}
= (\nu_n)_{l_1~\!\! l_2~\!\! \cdots~\!\! l_{n-1}~\!\! l_{n+1}} 
+ (\nu_n)_{l_1~\!\! l_2~\!\! \cdots~\!\! l_{n-2}~\!\! l_{n+1}~\!\! l_n}
+ \cdots 
+ (\nu_n)_{l_{n+1}~\!\! l_2~\!\! \cdots~\!\! l_n}.
\label{nu-Ritz-n}
\end{eqnarray}

Next, we derive an equation of motion for quantities with $n$ indices
based on the condition \eqref{nu-N-qu-n-E}.
We consider a periodic motion with the period $T$.
Because $(\nu_n)_{l_1~\!\! l_2~\!\! \cdots~\!\! l_n}$ has $n$ indices,
we suppose that a dynamical variable containing it as a frequency also possesses $n$ indices.
Then, the quantities $(A_a)_{l_1~\!\! l_2~\!\! \cdots~\!\! l_n}(t)$ 
$(a=1, 2, \cdots, n)$ have a feature
that $(A_a)_{l_1~\!\! l_2~\!\! \cdots~\!\! l_n}(t+T)
=(A_a)_{l_1~\!\! l_2~\!\! \cdots~\!\! l_n}(t)$
and can be described as
\begin{eqnarray}
(A_a)_{l_1~\!\! l_2~\!\! \cdots~\!\! l_n}(t)
= (A_a)_{l_1~\!\! l_2~\!\! \cdots~\!\! l_n}(0) 
e^{2\pi i (\nu_n)_{l_1~\!\! l_2~\!\! \cdots~\!\! l_n}t},
\label{A-n}
\end{eqnarray}
where $(\nu_n)_{l_1~\!\! l_2~\!\! \cdots~\!\! l_n}$ follows
the condition \eqref{nu-N-qu-n-E} and is also written by
$(\nu_n)_{l_1~\!\! l_2~\!\! \cdots~\!\! l_n} = n_{l_1~\!\! l_2~\!\! \cdots~\!\! l_n}/T$
($n_{l_1~\!\! l_2~\!\! \cdots~\!\! l_n}$ : an integer).

If we require that the product $(A_1A_2\cdots A_n)_{l_1~\!\! l_2~\!\! \cdots~\!\! l_n}(t)$ 
should take the same form as eq.~\eqref{A-n}, i.e.,
$(A_1A_2\cdots A_n)_{l_1~\!\! l_2~\!\! \cdots~\!\! l_n}(t)
=(A_1A_2\cdots A_n)_{l_1~\!\! l_2~\!\! \cdots~\!\! l_n}(0) 
e^{2\pi i (\nu_n)_{l_1~\!\! l_2~\!\! \cdots~\!\! l_n}t}$,
$(A_a)_{l_1~\!\! l_2~\!\! \cdots~\!\! l_n}(t)$ 
become components of extended matrices whose product is defined by\footnote{
This definition of the $n$-fold product is the same as that in Ref.~\cite{Xiong}.}
\begin{align}
(A_1A_2\cdots A_n)_{l_1~\!\! l_2~\!\! \cdots~\!\! l_n}(t)
\equiv \sum_{l_{n+1}} 
(A_1)_{l_1~\!\! l_2~\!\! \cdots~\!\! l_{n-1}~\!\! l_{n+1}}(t)
(A_2)_{l_1~\!\! l_2~\!\! \cdots~\!\! l_{n+1}~\!\! l_n}(t)
\cdots (A_n)_{l_{n+1}~\!\! l_2~\!\! \cdots~\!\! l_n}(t),
\label{A1A2...An}
\end{align}
where we use eq.~\eqref{nu-Ritz-n}.
We refer to $n(\ge 3)$-index variables,
which follow the product rule \eqref{A1A2...An}, as generalized matrices
or $n$-th power matrices.

Because the components with $l_{r} = l_{r'}$ on 
$(A_a)_{l_1~\!\! l_2~\!\! \cdots~\!\! l_n}(t)$ do not depend on time
as seen from eqs.~\eqref{nu-N-qu-n} and \eqref{A-n}, any conserved quantity 
satisfying 
$dC_{l_1~\!\! l_2~\!\! \cdots~\!\! l_n}/dt = 0$ can be described as 
a generalized matrix whose non-vanishing components are given by
\begin{eqnarray}
C_{l_1~\!\! \cdots~\!\! l_i~\!\! \cdots~\!\! 
l_{j-1}~\!\! l_{i}~\!\! l_{j+1}~\!\! \cdots~\!\! l_n} 
= \sum_{u (\ne i, j)} c_{l_{u}~\!\! l_{i}},
\label{A(N)-n}
\end{eqnarray}
where $c_{l_{u}~\!\! l_{i}}$ are constants, $i$ and $j(\ne i)$ 
are some numbers from $1$ to $n$,
$u$ runs from 1 to $n$ except for $i$ and $j$,
and the $n-1$ indices 
$(l_1, \cdots, l_i, \cdots l_{j-1}, l_{j+1}, \cdots, l_n)$ are all different.
We note that other components are zero.
We refer to this type of generalized matrix as a normal form,
a normal generalized matrix or an $n$-th power normal matrix.

Then, in a closed system, the Hamiltonians $(H_{a_k})_{l_1~\!\! l_2~\!\! \cdots~\!\! l_n}$ 
($k=1, 2 \cdots, n-1$) conserve and take the normal forms
whose non-vanishing components are given by
\begin{eqnarray}
(H_{a_k})_{l_1~\!\! \cdots~\!\! l_i~\!\! \cdots~\!\! 
l_{j-1}~\!\! l_{i}~\!\! l_{j+1}~\!\! \cdots~\!\! l_n} 
= \sum_{u (\ne i, j)} (h_{a_k})_{l_{u}~\!\! l_{i}},
\label{H-n}
\end{eqnarray}
where $(h_{a_k})_{l_{u}~\!\! l_{i}}$ are constants, $i$ and $j(\ne i)$ 
are some numbers from $1$ to $n$,
$u$ runs from 1 to $n$ except for $i$ and $j$,
and the $n-1$ indices 
$(l_1, \cdots, l_i, \cdots l_{j-1}, l_{j+1}, \cdots, l_n)$ are all different.

Using eqs.~\eqref{nu-N-qu-n-E}, \eqref{nu-Ritz-n}, \eqref{A-n}, 
\eqref{A1A2...An} and \eqref{H-n}, 
we derive the equation:
\begin{align}
\frac{d}{dt} A_{l_1~\!\! l_2~\!\! \cdots~\!\! l_n}(t) 
&= 2\pi i (\nu_n)_{l_1~\!\! l_2~\!\! \cdots~\!\! l_n}A_{l_1~\!\! l_2~\!\! \cdots~\!\! l_n}(t)
\nonumber \\
&= 2\pi i \beta \sum_{k=1}^{n} (-1)^{n-k} \sum_{(a_1, a_2, \cdots, a_{n-1})}{\rm sgn(P)}
\nonumber \\
&~~~~ \times 
\prod_{r=1}^{k-1}(E_{a_r})_{l_r~\!\! l_k}\prod_{r'=k}^{n-1}(E_{a_{r'}})_{l_{r'+1}~\!\! l_k}
A_{l_1~\!\! l_2~\!\! \cdots~\!\! l_n}(t)
\nonumber \\
&= \frac{1}{i \hbar} 
\left[A(t), H_1, H_2, \cdots, H_{n-1}\right]_{l_1~\!\! l_2~\!\! \cdots~\!\! l_n},
\label{gH-eq-n-derivation}
\end{align}
by choosing $\beta$, $(E_{a_r})_{l_{r}~\!\! l_{k}}$ and $(E_{a_{r'}})_{l_{r'+1}~\!\! l_{k}}$ as
\begin{eqnarray}
&~& \beta = \frac{\gamma}{h}(-1)^{\frac{(n-2)(n-3)}{2}+1},~~
\label{beta-n}\\
&~& (E_{a_r})_{l_{r}~\!\! l_{k}} = (h_{a_r})_{l_{r}~\!\! l_{k}}~~(r=1, \cdots, k-1),~~
\label{E-h-r}\\
&~& (E_{a_{r'}})_{l_{r'+1}~\!\! l_{k}} = (h_{a_{r'}})_{l_{r'+1}~\!\! l_{k}}~~(r'=k, \cdots, n-1),
\label{E-h-r'}
\end{eqnarray}
where $\gamma = 1$ for an odd number $n(\ge 3)$ and $\gamma = n-2$ for an even number $n(\ge 4)$.
Using eqs.~\eqref{nu-N-qu-n-E}, \eqref{beta-n}, 
\eqref{E-h-r} and \eqref{E-h-r'}, the frequency is represented by
\begin{align}
(\nu_n)_{l_1~\!\! l_2~\!\! \cdots l_n} 
&= \frac{\gamma}{h}(-1)^{\frac{(n-2)(n-3)}{2}+1}
\sum_{k=1}^{n} (-1)^{n-k} \sum_{(a_1, a_2, \cdots, a_{n-1})}{\rm sgn(P)} 
\nonumber \\
&~~~~ \times
\prod_{r=1}^{k-1}(h_{a_r})_{l_r~\!\! l_k}\prod_{r'=k}^{n-1}(h_{a_{r'}})_{l_{r'+1}~\!\! l_k}.
\label{nu-N-qu-n-h}
\end{align}
In eq.~\eqref{gH-eq-n-derivation}, 
$\left[A(t), H_1, H_2, \cdots, H_{n-1}\right]_{l_1~\!\! l_2~\!\! \cdots~\!\! l_n}$
is the $n$-fold commutator defined by
\begin{eqnarray}
\hspace{-1cm}
&~& \left[A(t), H_1, H_2, \cdots, H_{n-1}\right]_{l_1~\!\! l_2~\!\! \cdots~\!\! l_n}
\nonumber \\
\hspace{-1cm}
&~& ~~~~ \equiv \sum_{(a_1, a_2, \cdots, a_{n-1})}{\rm sgn(P)}
\sum_{l_{n+1}} \left\{A(t)_{l_1~\!\! l_2~\!\! \cdots~\!\! l_{n-1}~\!\! l_{n+1}} 
(H_{a_1})_{l_1~\!\! l_2~\!\! \cdots~\!\! l_{n-2}~\!\! l_{n+1}~\!\! l_n}
(H_{a_2})_{l_1~\!\! l_2~\!\! \cdots~\!\! l_{n-3}~\!\! l_{n+1}~\!\! l_{n-1}~\!\! l_n}\right.
\nonumber \\
&~& ~~~~~~~ \cdots
(H_{a_{n-1}})_{l_{n+1}~\!\! l_2~\!\! \cdots~\!\! l_{n}}
- (H_{a_1})_{l_1~\!\! l_2~\!\! \cdots~\!\! l_{n-1}~\!\! l_{n+1}} 
A(t)_{l_1~\!\! l_2~\!\! \cdots~\!\! l_{n-2}~\!\! l_{n+1}~\!\! l_n}
(H_{a_2})_{l_1~\!\! l_2~\!\! \cdots~\!\! l_{n-3}~\!\! l_{n+1}~\!\! l_{n-1}~\!\! l_n}
\nonumber \\
&~& ~~~~~~~ \cdots
(H_{a_{n-1}})_{l_{n+1}~\!\! l_2~\!\! \cdots~\!\! l_{n}} + \cdots 
+ (-1)^{n-1} (H_{a_1})_{l_1~\!\! l_2~\!\! \cdots~\!\! l_{n-1}~\!\! l_{n+1}} 
(H_{a_2})_{l_1~\!\! l_2~\!\! \cdots~\!\! l_{n-2}~\!\! l_{n+1}~\!\! l_n}
\nonumber \\
&~& ~~~~~~~ \left. \times
(H_{a_3})_{l_1~\!\! l_2~\!\! \cdots~\!\! l_{n-3}~\!\! l_{n+1}~\!\! l_{n-1}~\!\! l_n}
\cdots
(H_{a_{n-1}})_{l_{1}~\!\! l_{n+1}~\!\! l_{3}~\!\! \cdots~\!\! l_{n}}
A(t)_{l_{n+1}~\!\! l_2~\!\! \cdots~\!\! l_{n}}\right\}.
\label{n-pleC}
\end{eqnarray}

In this way, we have arrived at a generalization of Heisenberg's equation of motion:
\begin{eqnarray}
\frac{d}{dt} A_{l_1~\!\! l_2~\!\! \cdots~\!\! l_n}(t)
= \frac{1}{i \hbar} 
\left[A(t), H_1, H_2, \cdots, H_{n-1}\right]_{l_1~\!\! l_2~\!\! \cdots~\!\! l_n}.
\label{gH-eq-n}
\end{eqnarray}
In the same way as the dynamics of cubic matrices,
it is reconfirmed that the components with identical indices
on $A_{l_1~\!\! l_2~\!\! \cdots~\!\! l_n}(t)$
are independent of time, using eq.~\eqref{gH-eq-n} and the relation
$[B, C_1, \cdots, C_{n-2}, C_{n-1}]_{l_1~\!\! l_2~\!\! \cdots~\!\! l_n}=0$ 
that holds concerning a generalized matrix $B$ 
whose components with all different indices vanish, i.e., 
$B_{l_1~\!\! l_2~\!\! \cdots~\!\! l_n}=0$ ($l_i \ne l_j$ for all $i$ and $j$)
and arbitrary normal cubic matrices
$C_a$ $(a=1, 2, \cdots, n-1)$.
The validity of $[B, C_1, \cdots, C_{n-2}, C_{n-1}]_{l_1~\!\! l_2~\!\! \cdots~\!\! l_n}=0$ 
is understood from the following reason.
For $B$ and $C_a$, we obtain the relation:
\begin{align}
(BC_1\cdots C_{n-2}C_{n-1})_{l_1~\!\! l_2~\!\! \cdots~\!\! l_n}
&= \sum_{l_{n+1}} 
B_{l_1~\!\! l_2~\!\! \cdots~\!\! l_{n-1}~\!\! l_{n+1}}
(C_1)_{l_1~\!\! l_2~\!\! \cdots~\!\! l_{n-2}~\!\! l_{n+1}~\!\! l_n} \cdots 
\nonumber \\
&~~~~~ \times (C_{n-2})_{l_1~\!\! l_{n+1}~\!\! l_3~\!\! \cdots~\!\! l_n}
(C_{n-1})_{l_{n+1}~\!\! l_2~\!\! \cdots~\!\! l_n}=0
\label{C1C2...Cn}
\end{align}
from the fact that there exists no instance in which
each of indices 
$(l_1, l_2, \cdots, l_{n-2}, l_{n+1}, l_n)$, $\cdots$,
$(l_1, l_{n+1}, l_3, \cdots, l_n)$
and $(l_{n+1}, l_2, \cdots, l_n)$ contains exactly two identical indices,
other indices are all different, $(l_1, l_2, \cdots, l_{n-1}, l_{n+1})$
contains two or more than two identical indices,
and furthermore, the identical index is shared between them.
The equation~\eqref{gH-eq-n} coincides with the generalized Heisenberg equation 
obtained in Ref.~\cite{YK4} by employing the generalized Ritz rule 
and the generalized Bohr frequency condition. 
However, the generalized Bohr frequency condition used there does not 
have a clearly presented classical counterpart.

Here, we give comments on the equation~\eqref{gH-eq-n}.

One is about a symmetry in the equation.
Eq.~\eqref{gH-eq-n} is invariant under the transformation
of Hamiltonians $(H_{a_k})_{l_1~\!\! l_2~\!\! \cdots~\!\! l_n}
\to (H'_{a_k})_{l_1~\!\! l_2~\!\! \cdots~\!\! l_n}$
whose values given by eq.~\eqref{H-n} are transformed as
\begin{eqnarray}
(h_{a_k})_{l_{u}~\!\! l_{i}} \to (h'_{a_k})_{l_{u}~\!\! l_{i}}
 = (h_{a_k})_{l_{u}~\!\! l_{i}} + c_{a_k},
\label{H-inv}
\end{eqnarray}
where $c_{a_k}$ are constants.
Actually, it is shown that the following relation holds on,
\begin{eqnarray}
\left[A(t), H'_1, H'_2, \cdots, H'_{n-1}\right]_{l_1~\!\! l_2~\!\! \cdots~\!\! l_n}
= \left[A(t), H_1, H_2, \cdots, H_{n-1}\right]_{l_1~\!\! l_2~\!\! \cdots~\!\! l_n}
\label{equality}
\end{eqnarray}
using the relations:
\begin{align}
\left[A(t), H_{a_1}, H_{a_2}, 
\cdots, H_{a_{n-2}}, I\right]_{l_1~\!\! l_2~\!\! \cdots~\!\! l_n} 
&= - h (\delta \nu'_{n-1})_{l_1~\!\! l_2~\!\! \cdots~\!\! l_n}
A_{l_1~\!\! l_2~\!\! \cdots~\!\! l_n}(t)
\nonumber \\
&= -h \sum_{i=1}^{n}(-1)^{i+1}(\nu'_{n-1})_{l_1~\!\! l_2~\!\! \cdots~\!\! 
\hat{l}_i~\!\! \cdots l_n}
A_{l_1~\!\! l_2~\!\! \cdots~\!\! l_n}(t)
= 0,
\label{AHI'}\\
\left[A(t), H_{a_1}, \cdots, H_{a_{k}}, I, \cdots, I\right]_{l_1~\!\! l_2~\!\! \cdots~\!\! l_n} 
&= 0 ~~~~ (k=1, \cdots, n-3),
\label{AHII}\\
\left[A(t), I, \cdots, I\right]_{l_1~\!\! l_2~\!\! \cdots~\!\! l_n} &= 0,
\label{AII}
\end{align}
where $I$ is a specific type of normal generalized matrix whose non-vanishing components
are given by
\begin{eqnarray}
I_{l_1~\!\! \cdots~\!\! l_i~\!\! \cdots~\!\! 
l_{j-1}~\!\! l_{i}~\!\! l_{j+1}~\!\! \cdots~\!\! l_n} = 1,
\label{I-n}
\end{eqnarray}
and eqs.\eqref{AHII} and \eqref{AII} are understood from the definition of
the $n$-fold commutator.
In eq.~\eqref{I-n}, $i$ and $j(\ne i)$ are some numbers from $1$ to $n$,
and $(l_1, \cdots, l_i, \cdots l_{j-1}, l_{j+1}, \cdots, l_n)$ are all different.
In eq.~\eqref{AHI'},
$(\nu'_{n-1})_{l_1~\!\! l_2~\!\! \cdots l_{n-1}}$ is given by
\begin{align}
(\nu'_{n-1})_{l_1~\!\! l_2~\!\! \cdots l_{n-1}} 
&= \frac{\gamma}{h}(-1)^{\frac{(n-3)(n-4)}{2}+1}
\sum_{k=1}^{n-1} (-1)^{n-1-k} \sum_{(a_1, a_2, \cdots, a_{n-2})}{\rm sgn(P)} 
\nonumber \\
&~~~~ \times
\prod_{r=1}^{k-1}(h_{a_r})_{l_r~\!\! l_k}\prod_{r'=k}^{n-2}(h_{a_{r'}})_{l_{r'+1}~\!\! l_k}
\label{nu-n-1}
\end{align}
with $\gamma = 1$ for an odd number $n(\ge 3)$ and $\gamma = n-2$ 
for an even number $n(\ge 4)$.
In eq.~\eqref{AHI'},
the combination rule $(h_{a_k})_{l_{s}~\!\! l_{t}} 
= (h_{a_k})_{l_{s}~\!\! l_{u}} + (h_{a_k})_{l_{u}~\!\! l_{t}}
= (h_{a_k})_{l_{s}~\!\! l_{u}} - (h_{a_k})_{l_{t}~\!\! l_{u}}$ is used
to show $(\delta \nu'_{n-1})_{l_1~\!\! l_2~\!\! \cdots~\!\! l_n}=0$.
We notice that $I$ behaves like the identity,
because the following relation holds
\begin{align}
(AI\cdots I)_{l_1~\!\! l_2~\!\! \cdots~\!\! l_n}
= (IAI\cdots I)_{l_1~\!\! l_2~\!\! \cdots~\!\! l_n}
= \cdots = (I\cdots IA)_{l_1~\!\! l_2~\!\! \cdots~\!\! l_n}
=A_{l_1~\!\! l_2~\!\! \cdots~\!\! l_n}
\label{AIII}
\end{align}
for the components $A_{l_1~\!\! l_2~\!\! \cdots~\!\! l_n}$
with all different indices.

The other is about a solution of eq.~\eqref{gH-eq-n}, 
in the case that eq.~\eqref{gH-eq-n} is regarded as a master equation.
There is another solution such that
\begin{eqnarray}
&~& A_{l_1~\!\! l_2~\!\! \cdots~\!\! l_n}(t)
= A_{l_1~\!\! l_2~\!\! \cdots~\!\! l_n}(0) 
e^{2\pi i (\tilde{\nu}_n)_{l_1~\!\! l_2~\!\! \cdots~\!\! l_n}t},
\label{A-an}\\
&~& (\tilde{\nu}_n)_{l_1~\!\! l_2~\!\! \cdots l_n} 
=(\delta \tilde{\nu}'_{n-1})_{l_1~\!\! l_2~\!\! \cdots l_n}
= \sum_{i=1}^{n}(-1)^{i+1}(\tilde{\nu}'_{n-1})_{l_1~\!\! l_2~\!\! \cdots~\!\! 
\hat{l}_i~\!\! \cdots l_n}, 
\label{nu-an}
\end{eqnarray}
where $(\tilde{\nu}'_{n-1})_{l_1~\!\! l_2~\!\! \cdots l_{n-1}}$ is given by
\begin{align}
(\tilde{\nu}'_{n-1})_{l_1~\!\! l_2~\!\! \cdots l_{n-1}} 
&= \frac{{\gamma}}{h}(-1)^{\frac{(n-3)(n-4)}{2}+1}
\sum_{k=1}^{n-1} (-1)^{n-1-k} \sum_{(a_1, a_2, \cdots, a_{n-2})}{\rm sgn(P)} 
\nonumber \\
&~~~~ \times
\prod_{r=1}^{k-1}(\tilde{h}_{a_r})_{l_r~\!\! l_k}
\prod_{r'=k}^{n-2}(\tilde{h}_{a_{r'}})_{l_{r'+1}~\!\! l_k}
\label{nu-n-1-an}
\end{align}
with ${\gamma} = 1$ for an odd number $n(\ge 3)$ and ${\gamma} = n-2$ 
for an even number $n(\ge 4)$.
In eq.~\eqref{nu-n-1-an}, 
$(\tilde{h}_{a_r})_{l_{s}~\!\! l_{t}}$ are anti-symmetric constants
but do not satisfy the combination rule, i.e.,
$(\tilde{h}_{a_k})_{l_{s}~\!\! l_{t}} 
\ne (\tilde{h}_{a_k})_{l_{s}~\!\! l_{u}} + (\tilde{h}_{a_k})_{l_{u}~\!\! l_{t}}$.
In this case, Hamiltonians are normal forms whose non-vanishing components are given by
\begin{eqnarray}
&~& (H_{a_k})_{l_1~\!\! \cdots~\!\! l_i~\!\! \cdots~\!\! 
l_{j-1}~\!\! l_{i}~\!\! l_{j+1}~\!\! \cdots~\!\! l_n} 
= \sum_{u (\ne i, j)} (\tilde{h}_{a_k})_{l_{u}~\!\! l_{i}}~~~~ (k=1, 2, \cdots, n-2),
\label{Ha(N)-an}\\
&~& (H_{n-1})_{l_1~\!\! \cdots~\!\! l_i~\!\! \cdots~\!\! 
l_{j-1}~\!\! l_{i}~\!\! l_{j+1}~\!\! \cdots~\!\! l_n} = 1.
\label{Hn-2(N)-an}
\end{eqnarray}
From eq.~\eqref{nu-an}, we find that 
$(\tilde{\nu}_n)_{l_1~\!\! l_2~\!\! \cdots l_n}$
is $n$-coboundary and satisfies the cocycle condition as seen from the nilpotency
of coboundary operator, i.e., 
\begin{eqnarray}
(\delta \tilde{\nu}_n)_{l_1~\!\! l_2~\!\! \cdots l_{n+1}}
= (\delta^2 \tilde{\nu}'_{n-1})_{l_1~\!\! l_2~\!\! \cdots l_{n+1}} = 0.
\label{delta^2-an}
\end{eqnarray}

\subsection{Discretization of Jacobian}

Eq.~\eqref{gH-eq-n} resembles the Nambu equation
for $A=A(x_1, x_2, \cdots, x_n)$:
\begin{eqnarray}
\frac{dA}{dt} = \{A, H_1, H_2, \cdots, H_{n-1}\}_{\rm NB},
\label{N-eq-n}
\end{eqnarray}
where $x_i$ $(i=1,2, \cdots, n)$ are canonical $n$-plets,
$\{A, H_1, H_2, \cdots, H_{n-1}\}_{\rm NB}$ is the Nambu bracket 
and $\{A_1, A_2, A_3, \cdots, A_{n}\}_{\rm NB}$ is defined by
\begin{align}
\{A_1, A_2, A_3, \cdots, A_{n}\}_{\rm NB}
&\equiv \frac{\partial(A_1, A_2, A_3, \cdots, A_{n})}{\partial(x_1, x_2, x_3, \cdots, x_{n})}
\nonumber \\
&= \sum_{(a_1, a_2, \cdots, a_{n})}{\rm sgn(P)}
 \frac{\partial A_{a_1}}{\partial x_1} \frac{\partial A_{a_2}}{\partial x_2} 
\frac{\partial A_{a_3}}{\partial x_3} \cdots
 \frac{\partial A_{a_{n}}}{\partial x_{n}},
\label{Jacobian-n}
\end{align}
where sgn(P) is $+1$ and $-1$ for even and odd permutations among the set of subscripts
$(a_1, a_2, \cdots, a_{n})$, respectively, based on $+1$ 
for $(a_1, a_2, \cdots, a_{n})=(1, 2, \cdots, n)$,
and each $a_k$ $(k=1, 2, \cdots, n)$ runs from $1$ to $n$.
Hence, we conjecture that 
the generalized Heisenberg's equation of motion \eqref{gH-eq-n} 
can be a counterpart of the Nambu equation
and the multiple commutator defined by
\begin{eqnarray}
\hspace{-1cm}
&~& \left[A_1, A_2, \cdots, A_{n}\right]_{l_1~\!\! l_2~\!\! \cdots~\!\! l_n}
\nonumber \\
\hspace{-1cm}
&~& ~~~~ \equiv \sum_{(a_1, a_2, \cdots, a_{n})}{\rm sgn(P)}
\sum_{l_{n+1}} (A_{a_1})_{l_1~\!\! l_2~\!\! \cdots~\!\! l_{n-1}~\!\! l_{n+1}} 
(A_{a_2})_{l_1~\!\! l_2~\!\! \cdots~\!\! l_{n-2}~\!\! l_{n+1}~\!\! l_n}
\cdots
(A_{a_n})_{l_{n+1}~\!\! l_2~\!\! \cdots~\!\! l_{n}}
\label{n-pleC-again}
\end{eqnarray}
can become a candidate of discrete or quantum version of the Nambu bracket or the Jacobian.

Let us study whether the Nambu bracket and the multiple commutator possess
common features or not.
The basic features of the Nambu bracket are listed as follows,
\begin{align}
& \{A_{a_1}, A_{a_2}, \cdots, A_{a_n}\}_{\rm NB}
= {\rm sgn(P)}\{A_{1}, A_{2}, \cdots, A_{n}\}_{\rm NB}~~~~~({\rm skew~symmetry}),
\label{NB-1}\\
& \{A_{1}+B_{1}, A_{2}, \cdots, A_{n}\}_{\rm NB}
 =  \{A_{1}, A_{2}, \cdots, A_{n}\}_{\rm NB}
 +  \{B_{1}, A_{2}, \cdots, A_{n}\}_{\rm NB}~~~~~({\rm linearity}),
\label{NB-2}\\
& \{\{A_{1}, A_{2}, \cdots, A_{n}\}_{\rm NB}, B_{1}, B_{2}, \cdots, B_{n-1}\}_{\rm NB}
\nonumber \\
& ~~~~~
= \sum_{i=1}^n \{A_{1}, A_{2}, \cdots, \{A_{i}, B_{1}, B_{2}, \cdots, B_{n-1}\}_{\rm NB}, 
\cdots, A_{n}\}_{\rm NB}
~~~~~({\rm fundamental~identity}),
\label{NB-3}\\
& \{A_{1}A_{2}\cdots A_{n}, B_{1}, B_{2}, \cdots, B_{n-1}\}_{\rm NB}
 = \sum_{i=1}^n A_{1} A_{2} \cdots \{A_{i}, B_{1}, B_{2}, \cdots, B_{n-1}\}_{\rm NB}
\cdots A_{n}
\nonumber \\
&~~~~~~~~~~~~~~~~~~~~~~~~~~~~~~~~~~~~~~~~~~~~~~~~~~~~~~~~
~~~~~~~~~~~~~~~~~~~~~~~~~~~~~~~~~~~~~~~~~~~~~~~~~~~~~~~~~({\rm derivation~rule}).
\label{NB-4}
\end{align}

The multiple commutator of generalized matrices has the skew symmetry 
and the linearity, but does not necessarily satisfy the fundamental identity 
and the derivation rule. 
When $B_j$ $(j=1, 2, \cdots, n-1)$ are normal generalized matrices
and the eigenvalues $(f_n)_{l_{1}~\!\! l_2~\!\! \cdots~\!\! l_{n}}$ satisfy
the cocycle condition $(\delta f_n)_{l_{1}~\!\! l_2~\!\! \cdots~\!\! l_{n+1}}=0$,
we find that the multiple commutator of generalized matrices follows
the fundamental identity and the derivation rule, in the following way.
Here, the eigenvalues $(f_n)_{l_{1}~\!\! l_2~\!\! \cdots~\!\! l_{n}}$ are
specified by the eigenvalue equation:
\begin{eqnarray}
\left[A, B_1, B_2, \cdots, B_{n-1}\right]_{l_1~\!\! l_2~\!\! \cdots~\!\! l_n}
= (f_n)_{l_{1}~\!\! l_2~\!\! \cdots~\!\! l_{n}} A_{l_{1}~\!\! l_2~\!\! \cdots~\!\! l_{n}}
\label{eigen}
\end{eqnarray}
and the cocycle condition is denoted by
\begin{align}
(\delta f_n)_{l_1~\!\! l_2~\!\! \cdots~\!\! l_{n+1}}
= (-1)^{n+1} \left\{(f_n)_{l_1~\!\! l_2~\!\! \cdots~\!\! l_n}
- \sum_{i=1}^{n} (-1)^{n-i} 
(f_n)_{l_1~\!\! \cdots~\!\! \hat{l}_i~\!\! \cdots~\!\! l_{n}~\!\! l_{n+1}}\right\} = 0.
\label{cocycle-f}
\end{align}
Using eq.~\eqref{eigen}, the left-hand side and the right-hand side of 
the fundamental identity can be calculated as
\begin{align}
&\left[[A_1, A_2, \cdots, A_n], 
B_1, B_2, \cdots, B_{n-1}\right]_{l_1~\!\! l_2~\!\! \cdots~\!\! l_n}
= (f_n)_{l_{1}~\!\! l_2~\!\! \cdots~\!\! l_{n}} 
[A_1, A_2, \cdots, A_n]_{l_{1}~\!\! l_2~\!\! \cdots~\!\! l_{n}},
\label{lhs-comm}\\
&\sum_{i=1}^n \left[A_{1}, \cdots, [A_{i}, B_{1}, B_{2}, \cdots, B_{n-1}], 
\cdots, A_{n}\right]
\nonumber \\
&~~~~~~~~~~~~~~~~~~~ = \sum_{i=1}^n 
(f_n)_{l_1~\!\! \cdots~\!\! l_{i-1}~\!\! l_{n+1}~\!\! l_{i+1}~\!\! \cdots~\!\! l_{n}}
\left[A_{1}, A_{2}, \cdots, A_{n}\right]_{l_1~\!\! l_2~\!\! \cdots~\!\! l_n}
\nonumber \\
&~~~~~~~~~~~~~~~~~~~ = \sum_{i=1}^n (-1)^{n-i} 
(f_n)_{l_1~\!\! \cdots~\!\! \hat{l}_i~\!\! \cdots~\!\! l_{n}~\!\! l_{n+1}}
\left[A_{1}, A_{2}, \cdots, A_{n}\right]_{l_1~\!\! l_2~\!\! \cdots~\!\! l_n},
\end{align}
respectively, and these agree with if the condition \eqref{cocycle-f} is met.
The same holds on for the derivation rule.

The properties of the fundamental identity and the derivation rule 
in the generalized matrix mechanics suggest that the quantum version 
of the generator of generalized canonical transformations in Nambu mechanics 
may be restricted to conserved quantities. 
Therefore, in Nambu mechanical systems 
where transformations generated by time-dependent quantities 
play an important role, it may be challenging to properly 
define their quantum counterparts.

\section{Conclusions and discussions}

Inspired by the fact that Heisenberg's matrix mechanics 
was derived from Hamiltonian mechanics under the guidance of the correspondence principle, 
we have investigated a mechanics involving discrete variables, 
starting from Nambu mechanics. 
As a result, we reconstructed a framework of mechanics for dynamical variables 
expressed in terms of generalized matrices, 
which can be regarded as an extension of matrix mechanics. 
The fundamental equation of this mechanics corresponds to 
a generalized version of Heisenberg's equation of motion, 
formulated using an $n$-fold commutator. 
This equation can also be interpreted as the counterpart to the Nambu equation 
associated with canonical $n$-plets. 
Based on this interpretation, we reconfirmed that the multiple commutator 
involving generalized matrices can serve as a quantum or discrete version 
of the Nambu bracket, namely the Jacobian.

In what follows, we discuss the generalized matrix mechanics 
and possible applications of the generalized matrices.

In our generalized dynamics, the objective is to determine
the values of Hamiltonians, i.e., $(h_{a_k})_{l_u~\!\! l_i}$ in eq.~\eqref{H-n},
taking the equation \eqref{gH-eq-n} as the master equation.
The relationship between symmetries and conservation laws holds
in a manner analogous to conventional matrix mechanics.
That is, generalized matrices $C^a$ that commute with the Hamiltonians, i.e.,
$[C^a, H_1, \cdots, H_{n-1}] = 0$, are identified as conserved quantities, 
as understood from the equation \eqref{gH-eq-n}.
Furthermore, arbitrary constant normal matrices become conserved quantities
because the relation $[N_1, \cdots, N_{n}]=0$ holds
for any $n$-th power normal matrices $N_b$ ($b=1, \cdots, n$).

In quantum mechanics, the matrix elements in matrix mechanics are 
interpreted as probability amplitudes representing transitions between quantum states. 
At present, it is unclear whether the generalized matrix elements 
with $n$ indices admit a similar interpretation, 
or whether the extended version of matrix mechanics 
can be applied to real physical systems,
although the generalized cubic matrix mechanics can describe
physical systems in the conventional matrix mechanics, as shown in the Appendix B. 
This suggests the need for further investigation into 
the understanding and application of generalized matrices, 
beyond the framework of a mere extension of Heisenberg's matrix mechanics 
or a counterpart to Nambu mechanics.

Next, regarding the use of generalized matrices
 and the discrete version of the Nambu bracket expressed via the multiple commutator: 
in quantum mechanics, spin variables are represented using matrices. 
Therefore, if an extended version of spin exists, 
it is expected to be representable using generalized matrices. 
In fact, in Ref.~\cite{YK5}, 
a generalization of spin algebra using generalized matrices has been proposed. 
Furthermore, in that work, uncertainty relations for spacetime 
have been derived by using the triple commutator involving spacetime coordinates 
expressed via cubic matrices. 
It is also conceivable that uncertainty relations 
for higher-dimensional spacetime can be derived using the $n$-fold commutator.

Finally, generalized matrices may prove useful 
in formulating the quantum theory of extended objects. 
The action integral in string theory includes the area of the worldsheet, 
and this area is expressed using a two-dimensional Jacobian~\cite{Schild,Yoneya2}. 
Similarly, $p$-dimensional extended objects 
are described using $(p+1)$-dimensional Jacobians. 
By discretizing the Jacobian, it may become possible 
to microscopically formulate various branes including membrane~\cite{deWH&N}, 
which could, in turn, be applicable to the formulation of M-theory\footnote{
Using Lie 3-algebra~\cite{Filippov}, an effective theory for multiple M2 branes
called the BLG model has been constructed~\cite{B&L1,B&L2,B&L3,Gustavsson}.
See Ref.~\cite{deA&I} for extensive reviews of $n$-nary algebra
and its applications.}.
Recently, there have been attempts relating to many-index objects,
for instance, to regularize the Nambu brackets using $2n$-index objects~\cite{AS&T}
and to construct gauge theories 
using a new type of non-associative algebra that includes 
both ordinary matrices and cubic matrices~\cite{BP&R}. 
These developments suggest that 
there may still be room for various extensions and applications.

\section*{Acknowledgments}
This work was supported in part by scientific grants 
from the Ministry of Education, Culture,
Sports, Science and Technology under Grant No.~22K03632.

\appendix

\section{Details of derivation}

In this appendix, details of derivation of eq.~\eqref{nu-N-qu-rel-n},
$(\delta\nu_n^{(0)})_{l_1~\!\! l_2~\!\! \cdots~\!\! l_n~\!\! l_{n+1}}=0$
and so on are provided, and 
it is shown that variables with $n$ totally anti-symmetric indices
satisfy the generalized Heisenberg equation \eqref{gH-eq-n}.

Starting from eq.~\eqref{nu-N-qu-0-n},
eq.~\eqref{nu-N-qu-rel-n} is derived as follows,
\begin{align}
(\nu_n^{(0)})_{l_1~\!\! l_2~\!\! \cdots~\!\! l_n}
&= {\beta}  \sum_{(a_1, a_2, \cdots, a_{n-1})}{\rm sgn(P)} 
\prod_{r=1}^{n-1}(E_{a_r})_{l_r~\!\! l_n}
\nonumber \\
&= {\beta}  \sum_{(a_1, a_2, \cdots, a_{n-1})}{\rm sgn(P)} 
(E_{a_1})_{l_1~\!\! l_n}(E_{a_2})_{l_2~\!\! l_n} \cdots (E_{a_{n-1}})_{l_{n-1}~\!\! l_n}
\nonumber \\
&= {\beta}  \sum_{(a_1, a_2, \cdots, a_{n-1})}{\rm sgn(P)} 
((E_{a_1})_{l_1~\!\! l_{n+1}}-(E_{a_1})_{l_{n}~\!\! l_{n+1}})
((E_{a_2})_{l_2~\!\! l_{n+1}}-(E_{a_2})_{l_{n}~\!\! l_{n+1}}) 
\nonumber \\
&~~~~ \times \cdots \times
((E_{a_{n-1}})_{l_{n-1}~\!\! l_{n+1}}-(E_{a_{n-1}})_{l_{n}~\!\! l_{n+1}})
\nonumber \\
& = {\beta}  \sum_{(a_1, a_2, \cdots, a_{n-1})}{\rm sgn(P)} 
\left\{(E_{a_1})_{l_1~\!\! l_{n+1}}(E_{a_2})_{l_2~\!\! l_{n+1}} 
\cdots (E_{a_{n-1}})_{l_{n-1}~\!\! l_{n+1}} \right.
\nonumber \\
&~~~~ - (E_{a_1})_{l_1~\!\! l_{n+1}}(E_{a_2})_{l_2~\!\! l_{n+1}} 
\cdots (E_{a_{n-2}})_{l_{n-2}~\!\! l_{n+1}} (E_{a_{n-1}})_{l_{n}~\!\! l_{n+1}}
\nonumber \\
&~~~~ - (E_{a_1})_{l_1~\!\! l_{n+1}}(E_{a_2})_{l_2~\!\! l_{n+1}} 
\cdots (E_{a_{n-2}})_{l_{n}~\!\! l_{n+1}} (E_{a_{n-1}})_{l_{n-1}~\!\! l_{n+1}}
\nonumber \\
&~~~~ \left. - \cdots - (E_{a_1})_{l_n~\!\! l_{n+1}}(E_{a_2})_{l_2~\!\! l_{n+1}} 
\cdots (E_{a_{n-2}})_{l_{n-2}~\!\! l_{n+1}} (E_{a_{n-1}})_{l_{n-1}~\!\! l_{n+1}}\right\}
\nonumber \\
& = (\nu_n^{(0)})_{l_1~\!\! l_2~\!\! \cdots~\!\! l_{n-1}~\!\! l_{n+1}}
- \sum_{i = 1}^{n-1}  
(\nu_n^{(0)})_{l_1~\!\! \cdots l_{i-1}~\!\! l_{n}~\!\!
l_{i+1}~\!\! \cdots~\!\! l_{n-1}~\!\! l_{n+1}}
\nonumber \\
& = (\nu_n^{(0)})_{l_1~\!\! l_2~\!\! \cdots~\!\! l_{n-1}~\!\! l_{n+1}}
- \sum_{i = 1}^{n-1} (-1)^{n-1-i}
(\nu_n^{(0)})_{l_1~\!\! \cdots \hat{l}_{i}~\!\! \cdots~\!\! l_{n}~\!\! l_{n+1}}
\nonumber \\
& = \sum_{i = 1}^n (-1)^{n-i} 
(\nu_n^{(0)})_{l_1~\!\! \cdots~\!\! \hat{l}_{i}~\!\! \cdots~\!\! l_{n}~\!\! l_{n+1}},
\label{A:nu-N-qu-rel-n-derivation}
\end{align}
where we use eqs.~\eqref{anti-sym-n} and \eqref{R-E-n} and
the feature that terms including both $(E_{a_{r}})_{l_{n}~\!\! l_{n+1}}$
and $(E_{a_{r'}})_{l_{n}~\!\! l_{n+1}}$ ($a_r \ne a_{r'}$) disappear 
because they are canceled out
due to totally antisymmetric property regarding $(a_1, a_2, \cdots, a_{n-1})$. 

For reference, the frequencies
$(\nu_n^{(0)})_{l_1~\!\! \cdots l_{i-1}~\!\! l_{n}~\!\!
l_{i+1}~\!\! \cdots~\!\! l_{n-1}~\!\! l_{n+1}}$
and $(\nu_n^{(0)})_{l_1~\!\! \cdots~\!\! \hat{l}_{i}~\!\! \cdots~\!\! l_{n}~\!\! l_{n+1}}$
are expressed as
\begin{align}
\hspace{-2cm}
&(\nu_n^{(0)})_{l_1~\!\! \cdots l_{i-1}~\!\! l_{n}~\!\!
l_{i+1}~\!\! \cdots~\!\! l_{n-1}~\!\! l_{n+1}}
 = {\beta}  \sum_{(a_1, a_2, \cdots, a_{n-1})}{\rm sgn(P)} 
(E_{a_1})_{l_1~\!\! l_{n+1}}(E_{a_2})_{l_2~\!\! l_{n+1}} \cdots
\nonumber \\
& ~~~~~~~~~~~~~~~~~~~~
 \times (E_{a_{i-1}})_{l_{i-1}~\!\! l_{n+1}}(E_{a_i})_{l_n~\!\! l_{n+1}}
(E_{a_{i+1}})_{l_{i+1}~\!\! l_{n+1}}
\cdots (E_{a_{n-1}})_{l_{n-1}~\!\! l_{n+1}},
\label{A:nu0-1}\\
\hspace{-2cm}
&(\nu_n^{(0)})_{l_1~\!\! \cdots~\!\! \hat{l}_{i}~\!\! \cdots~\!\! l_{n}~\!\! l_{n+1}}
= {\beta}  \sum_{(a_1, a_2, \cdots, a_{n-1})}{\rm sgn(P)} 
(E_{a_1})_{l_1~\!\! l_{n+1}}(E_{a_2})_{l_2~\!\! l_{n+1}} \cdots
\nonumber \\
& ~~~~~~~~~~~~~~~~~~~ 
\times (E_{a_{i-1}})_{l_{i-1}~\!\! l_{n+1}}(E_{a_i})_{l_{i+1}~\!\! l_{n+1}}
\cdots (E_{a_{n-2}})_{l_{n-1}~\!\! l_{n+1}} (E_{a_{n-1}})_{l_{n}~\!\! l_{n+1}},
\label{A:nu0-2}
\end{align}
respectively.

It is shown that $(\nu_n^{(0)})_{l_1~\!\! l_2~\!\! \cdots~\!\! l_n}$ satisfies
the cocycle condition
$(\delta\nu_n^{(0)})_{l_1~\!\! l_2~\!\! \cdots~\!\! l_n~\!\! l_{n+1}}=0$ as follows,
\begin{align}
(\delta\nu_n^{(0)})_{l_1~\!\! l_2~\!\! \cdots~\!\! l_{n+1}}
&= \sum_{i=1}^{n+1} (-1)^{i+1} 
(\nu_n^{(0)})_{l_1~\!\! \cdots~\!\! \hat{l}_i~\!\! \cdots~\!\! l_{n+1}}
= (-1)^n \sum_{i=1}^{n+1} (-1)^{n-i-1} 
(\nu_n^{(0)})_{l_1~\!\! \cdots~\!\! \hat{l}_i~\!\! \cdots~\!\! l_{n+1}}
\nonumber \\
&= (-1)^n \left\{-\sum_{i=1}^{n} (-1)^{n-i} 
(\nu_n^{(0)})_{l_1~\!\! \cdots~\!\! \hat{l}_i~\!\! \cdots~\!\! l_n~\!\! l_{n+1}}
+ (\nu_n^{(0)})_{l_1~\!\! l_2~\!\! \cdots~\!\! l_n}\right\} = 0,
\label{A:delta-n}
\end{align}
where we use eq.~\eqref{nu-N-qu-rel-n}.

It is shown that the frequency \eqref{nu-N-qu-n} 
have a cyclic property \eqref{nu-cyclic-n} as follows,
\begin{align}
(\nu_n)_{l_{p+1}~\!\! l_{p+2}~\!\! \cdots~\!\! l_{n}~\!\! l_1~\!\! \cdots~\!\! l_{p}}
&= \sum_{k=1}^{n} (-1)^{(n-1)k} 
(\nu_n^{(0)})_{l_{p+k+1}~\!\! l_{p+k+2}~\!\! \cdots~\!\! l_{n}~\!\! 
l_1~\!\! \cdots~\!\! l_{p+k}}
\nonumber \\
& 
= (-1)^{(n-1)p} \sum_{k=1}^{n} (-1)^{(n-1)(p+k)} 
(\nu_n^{(0)})_{l_{p+k+1}~\!\! l_{p+k+2}~\!\! \cdots~\!\! l_{n}~\!\! 
l_1~\!\! \cdots~\!\! l_{p+k}}
\nonumber \\
& 
= (-1)^{(n-1)p} \sum_{k=1}^{n} (-1)^{(n-1)k} 
(\nu_n^{(0)})_{l_{k+1}~\!\! l_{k+2}~\!\! \cdots~\!\! l_{n}~\!\! 
l_1~\!\! \cdots~\!\! l_{k}}
\nonumber \\
&
= (-1)^{(n-1)p} (\nu_n)_{l_1~\!\! l_2~\!\! \cdots~\!\! l_n}.
\label{A:nu-N-qu-n-proof}
\end{align}

Next, let us derive eq.~\eqref{nu-N-qu-n-E}.
Using eq.~\eqref{nu-N-qu-0-n}, we obtain the relation:
\begin{align}
(\nu_n^{(0)})_{l_{k+1}~\!\! l_{k+2}~\!\! \cdots~\!\! l_n~\!\! l_{1}~\!\! \cdots~\!\! l_{k}}
&= {\beta}\sum_{(a_1, a_2, \cdots, a_{n-1})}{\rm sgn(P)} 
\prod_{r=1}^{n-1}(E_{a_r})_{l_{r+k}~\!\! l_k}
\nonumber \\
&= {\beta}\sum_{(a_1, a_2, \cdots, a_{n-1})}{\rm sgn(P)} 
(E_{a_1})_{l_{k+1}~\!\! l_k}(E_{a_2})_{l_{k+2}~\!\! l_k}\cdots(E_{a_{n-k}})_{l_{n}~\!\! l_k}
\nonumber \\
& ~~~~ \times
(E_{a_{n-k+1}})_{l_{1}~\!\! l_k}(E_{a_{n-k+2}})_{l_{2}~\!\! l_k}
\cdots(E_{a_{n-1}})_{l_{k-1}~\!\! l_k}
\nonumber \\
& = {\beta}\sum_{(a_1, a_2, \cdots, a_{n-1})}{\rm sgn(P)} (-1)^{(n-k)(k-1)}
(E_{a_1})_{l_{1}~\!\! l_k}(E_{a_2})_{l_{2}~\!\! l_k}\cdots
\nonumber \\
& ~~~~ \times (E_{a_{k-1}})_{l_{k-1}~\!\! l_k}
(E_{a_{k}})_{l_{k+1}~\!\! l_k}(E_{a_{k+1}})_{l_{k+2}~\!\! l_k}
\cdots(E_{a_{n-1}})_{l_{n}~\!\! l_k}
\nonumber \\
& = {\beta} \sum_{(a_1, a_2, \cdots, a_{n-1})}{\rm sgn(P)}(-1)^{(n-k)(k-1)} 
\nonumber \\ 
&~~~~ \times
\prod_{r=1}^{k-1}(E_{a_r})_{l_r~\!\! l_k}\prod_{r'=k}^{n-1}(E_{a_{r'}})_{l_{r'+1}~\!\! l_k},
\label{A:nu-N-qu-n-E-derivation1}
\end{align}
where the factor $(-1)^{(n-k)(k-1)}$ appears from the $(n-k)(k-1)$ exchange of
the indices on $E_{a_r}$
from $(a_1, a_2, \cdots, a_{n-k}, a_{n-k+1}, a_{n-k+2}, \cdots, a_{n-1})$ to
$(a_{k}, a_{k+1}, \cdots, a_{n-1}, a_1, a_2, \cdots, a_{k-1})$.
By inserting eq.~\eqref{A:nu-N-qu-n-E-derivation1} into eq.~\eqref{nu-N-qu-n},
eq. \eqref{nu-N-qu-n-E} is obtained as follows,
\begin{align}
(\nu_n)_{l_1~\!\! l_2~\!\! \cdots l_n} 
&\equiv \sum_{k=1}^{n} (-1)^{(n-1)k} 
(\nu_n^{(0)})_{l_{k+1}~\!\! l_{k+2}~\!\! \cdots~\!\! l_{n}~\!\! l_1~\!\! \cdots~\!\! l_{k}}
\nonumber \\
&= \sum_{k=1}^{n} (-1)^{(n-1)k} {\beta} \sum_{(a_1, a_2, \cdots, a_{n-1})}{\rm sgn(P)}
(-1)^{(n-k)(k-1)}
\prod_{r=1}^{k-1}(E_{a_r})_{l_r~\!\! l_k}\prod_{r'=k}^{n-1}(E_{a_{r'}})_{l_{r'+1}~\!\! l_k},
\nonumber \\
&= {\beta} \sum_{k=1}^{n} (-1)^{n-k} \sum_{(a_1, a_2, \cdots, a_{n-1})}{\rm sgn(P)} 
\prod_{r=1}^{k-1}(E_{a_r})_{l_r~\!\! l_k}\prod_{r'=k}^{n-1}(E_{a_{r'}})_{l_{r'+1}~\!\! l_k},
\label{A:nu-N-qu-n-E-derivation2}
\end{align}
where we use $(n-1)k + (n-k)(k-1) = 2nk -k(k+1) - (n-k) = n-k$ (mod 2).
From the fact that eq.~\eqref{nu-N-qu-rel-n} holds for 
$(\nu_n)_{l_1~\!\! l_2~\!\! \cdots l_n}$, it is shown that
$(\nu_n)_{l_1~\!\! l_2~\!\! \cdots l_n}$ 
satisfies the rule \eqref{nu-Ritz-n} as follows,
\begin{align}
(\nu_n)_{l_1~\!\! l_2~\!\! \cdots~\!\! l_n}
&= \sum_{i=1}^n (-1)^{n-i} (\nu_n)_{l_1~\!\! \cdots~\!\! 
\hat{l}_i~\!\! \cdots~\!\! l_n~\!\! l_{n+1}}
= \sum_{i=1}^n (\nu_n)_{l_1~\!\! \cdots~\!\! l_{i-1}~\!\!
l_{n+1}~\!\! l_{i+1}~\!\! \cdots~\!\! l_n}
\nonumber \\
&= (\nu_n)_{l_{n+1}~\!\! l_2~\!\! \cdots~\!\! l_{n}} 
+ \cdots 
+ (\nu_n)_{l_1~\!\! l_2~\!\! \cdots~\!\! l_{n-2}~\!\! l_{n+1}~\!\! l_n}
+ (\nu_n)_{l_{1}~\!\! l_2~\!\! \cdots~\!\! l_{n-1}~\!\! l_{n+1}}
\nonumber \\
&= (\nu_n)_{l_1~\!\! l_2~\!\! \cdots~\!\! l_{n-1}~\!\! l_{n+1}} 
+ (\nu_n)_{l_1~\!\! l_2~\!\! \cdots~\!\! l_{n-2}~\!\! l_{n+1}~\!\! l_n}
+ \cdots 
+ (\nu_n)_{l_{n+1}~\!\! l_2~\!\! \cdots~\!\! l_n}.
\label{A-nu-Ritz-n}
\end{align}

Here, for completeness, we show that 
$(\nu_n^{(0)})_{l_1~\!\! l_2~\!\! \cdots~\!\! l_n}$ 
given by eq.~\eqref{nu-N-qu-0-n} have a cyclic property secretly.
Using eq.~\eqref{nu-N-qu-0-n}, we obtain the relation:
\begin{align}
(\nu_n^{(0)})_{l_{p+1}~\!\! l_{p+2}~\!\! \cdots~\!\! l_n~\!\! l_{1}~\!\! \cdots~\!\! l_{p}}
&= {\beta}\sum_{(a_1, a_2, \cdots, a_{n-1})}{\rm sgn(P)} 
\prod_{r=1}^{n-1}(E_{a_r})_{l_{r+p}~\!\! l_p}
\nonumber \\
&= {\beta}\sum_{(a_1, a_2, \cdots, a_{n-1})}{\rm sgn(P)} 
(E_{a_1})_{l_{p+1}~\!\! l_p}(E_{a_2})_{l_{p+2}~\!\! l_p}\cdots
(E_{a_{n-p-1}})_{l_{n-1}~\!\! l_p}(E_{a_{n-p}})_{l_{n}~\!\! l_p}
\nonumber \\
& ~~~~ \times
(E_{a_{n-p+1}})_{l_{1}~\!\! l_p}\cdots(E_{a_{n-1}})_{l_{p-1}~\!\! l_p}
\nonumber \\
&= {\beta}\sum_{(a_1, a_2, \cdots, a_{n-1})}{\rm sgn(P)} 
((E_{a_1})_{l_{p+1}~\!\! l_n}+(E_{a_1})_{l_{n}~\!\! l_p})
((E_{a_2})_{l_{p+2}~\!\! l_n}+(E_{a_2})_{l_{n}~\!\! l_p}) \cdots
\nonumber \\
& ~~~~ \times ((E_{a_{n-p-1}})_{l_{n-1}~\!\! l_n}+(E_{a_{n-p-1}})_{l_{n}~\!\! l_p})
(E_{a_{n-p}})_{l_{n}~\!\! l_p}
\nonumber \\
& ~~~~ \times ((E_{a_{n-p+1}})_{l_{1}~\!\! l_n}+(E_{a_{n-p+1}})_{l_{n}~\!\! l_p})
\cdots((E_{a_{n-1}})_{l_{p-1}~\!\! l_n}+(E_{a_{n-1}})_{l_{n}~\!\! l_p})
\nonumber \\
& = {\beta}\sum_{(a_1, a_2, \cdots, a_{n-1})}{\rm sgn(P)} 
(E_{a_1})_{l_{p+1}~\!\! l_n}(E_{a_2})_{l_{p+2}~\!\! l_n}\cdots
\nonumber \\
& ~~~~ \times (E_{a_{n-p-1}})_{l_{n-1}~\!\! l_n}
(-E_{a_{n-p}})_{l_{p}~\!\! l_n}(E_{a_{n-p+1}})_{l_{1}~\!\! l_n}
\cdots(E_{a_{n-1}})_{l_{p-1}~\!\! l_n}
\nonumber \\
& = -{\beta} \sum_{(a_1, a_2, \cdots, a_{n-1})}{\rm sgn(P)}
(E_{a_{n-p+1}})_{l_{1}~\!\! l_n}
\cdots(E_{a_{n-1}})_{l_{p-1}~\!\! l_n}
\nonumber \\
& ~~~~ \times (E_{a_{n-p}})_{l_{p}~\!\! l_n}
(E_{a_1})_{l_{p+1}~\!\! l_n}(E_{a_2})_{l_{p+2}~\!\! l_n}
\cdots(E_{a_{n-p-1}})_{l_{n-1}~\!\! l_n}
\nonumber \\
& = (-1)^{(n-1)p} {\beta} \sum_{(a_1, a_2, \cdots, a_{n-1})}{\rm sgn(P)}
(E_{a_{1}})_{l_{1}~\!\! l_n}
\cdots(E_{a_{p-1}})_{l_{p-1}~\!\! l_n}
\nonumber \\
& ~~~~ \times (E_{a_{p}})_{l_{p}~\!\! l_n}
(E_{a_{p+1}})_{l_{p+1}~\!\! l_n}(E_{a_{p+2}})_{l_{p+2}~\!\! l_n}
\cdots(E_{a_{n-1}})_{l_{n-1}~\!\! l_n}
\nonumber \\
& = (-1)^{(n-1)p}(\nu_n^{(0)})_{l_1~\!\! l_2~\!\! \cdots~\!\! l_n},
\label{A:nu-N-0-cyclic}
\end{align}
where the factor $-(-1)^{(n-1)p}$ comes from the $(n-2)(n-p-1) + (p-1)
(=(n-1)p +1 ({\rm mod}~2))$ exchange of
the indices on $E_{a_r}$
from $(a_{n-p+1}, \cdots, a_{n-1}, a_{n-p}, a_{1}, a_{2}, \cdots, a_{n-p-1})$ to
$(a_{1}, \cdots, a_{n-p-2}, a_{n-p-1}, a_{n-p}, a_{n-p+1}, \cdots, a_{n-1})$.  

Finally, let us show that a variable given by
\begin{eqnarray}
A_{l_1~\!\! l_2~\!\! \cdots~\!\! l_n}(t)
= A_{l_1~\!\! l_2~\!\! \cdots~\!\! l_n}(0)
e^{2\pi i (\nu_n)_{l_1~\!\! l_2~\!\! \cdots~\!\! l_n}t}
\label{A-A(t)}
\end{eqnarray}
with the frequency:
\begin{align}
(\nu_n)_{l_1~\!\! l_2~\!\! \cdots l_n} 
&= \frac{\gamma}{h}(-1)^{\frac{(n-2)(n-3)}{2}+1}
\sum_{k=1}^{n} (-1)^{n-k} \sum_{(a_1, a_2, \cdots, a_{n-1})}{\rm sgn(P)} 
\nonumber \\
&~~~~~ \times
\prod_{r=1}^{k-1}(h_{a_r})_{l_r~\!\! l_k}\prod_{r'=k}^{n-1}(h_{a_{r'}})_{l_{r'+1}~\!\! l_k}
\label{A-nu-N-qu-n-h}
\end{align}
satisfies the generalized Heisenberg equation:
\begin{eqnarray}
\frac{d}{dt} A_{l_1~\!\! l_2~\!\! \cdots~\!\! l_n}(t) 
= \frac{1}{i \hbar} 
\left[A(t), H_1, H_2, \cdots, H_{n-1}\right]_{l_1~\!\! l_2~\!\! \cdots~\!\! l_n},
\label{A-gH-eq-n}
\end{eqnarray}
where $(H_{a_k})_{l_1~\!\! l_2~\!\! \cdots~\!\! l_n}$ 
($k=1, 2 \cdots, n-1$) are the Hamiltonians with the normal forms
whose non-vanishing components are given by
\begin{eqnarray}
(H_{a_k})_{l_1~\!\! \cdots~\!\! l_i~\!\! \cdots~\!\! 
l_{j-1}~\!\! l_{i}~\!\! l_{j+1}~\!\! \cdots~\!\! l_n} 
= \sum_{u (\ne i, j)} (h_{a_k})_{l_{u}~\!\! l_{i}},
\label{A-H-n}
\end{eqnarray}
with constants $(h_{a_k})_{l_{u}~\!\! l_{i}}$. 
In eq.~\eqref{A-H-n}, $i$ and $j(\ne i)$ are some numbers from $1$ to $n$,
and $(l_1, \cdots, l_i, \cdots l_{j-1}, l_{j+1}, \cdots, l_n)$ are all different.
We notice that eq.~\eqref{A-nu-N-qu-n-h} agrees with eq.~\eqref{A:nu-N-qu-n-E-derivation2}
by identifying ${\beta}$, $(E_{a_r})_{l_{r}~\! l_{k}}$ 
and $(E_{a_{r'}})_{l_{r'+1}~\! l_{k}}$ as
\begin{eqnarray}
&~& {\beta} = \frac{\gamma}{h}(-1)^{\frac{(n-2)(n-3)}{2}+1},~~
\label{A-beta-n}\\
&~& (E_{a_r})_{l_{r}~\! l_{k}} = (h_{a_r})_{l_{r}~\! l_{k}}~~(r=1, \cdots, k-1),~~
\label{A-E-h-r}\\
&~& (E_{a_{r'}})_{l_{r'+1}~\! l_{k}} = (h_{a_{r'}})_{l_{r'+1}~\! l_{k}}~~(r'=k, \cdots, n-1),
\label{A-E-h-r'}
\end{eqnarray}
respectively.

First, exchanging the Hamiltonians, we obtain the relation:
\begin{eqnarray}
&~& \left[A(t), H_1, H_2, \cdots, H_{n-2}, H_{n-1}\right]_{l_1~\!\! l_2~\!\! \cdots~\!\! l_n}
\nonumber \\
&~& ~~~ = (-1)^{\frac{(n-2)(n-3)}{2}}
\left[A(t), H_{n-2}, \cdots, H_2, H_1, H_{n-1}\right]_{l_1~\!\! l_2~\!\! \cdots~\!\! l_n}.
\label{A-gH-rhs}
\end{eqnarray}

The first term $(A(t)H_{n-2}H_{n-3} \cdots H_2 H_1 H_{n-1})_{l_1~\!\! l_2~\!\! \cdots~\!\! l_n}$
in the right-hand side of eq.~\eqref{A-gH-rhs} is calculated as
\begin{align}
&(A(t)H_{n-2}H_{n-3} \cdots H_2 H_1 H_{n-1})_{l_1~\!\! l_2~\!\! \cdots~\!\! l_n}
\nonumber \\
&~~~= \sum_{l_{n+1}} A(t)_{l_1~\!\! l_2~\!\! \cdots~\!\! l_{n+1}}
(H_{n-2})_{l_1~\!\! \cdots~\!\! l_{n-2}~\!\! l_{n+1}~\!\! l_{n}}
(H_{n-3})_{l_1~\!\! \cdots~\!\! l_{n-3}~\!\! l_{n+1}~\!\! l_{n-1}~\!\! l_{n}}
\nonumber \\
&~~~~~~ \times \cdots \times
(H_{2})_{l_1~\!\! l_{2}~\!\! l_{n+1}~\!\! l_{4}~\!\! \cdots~\!\! l_{n}}
(H_{1})_{l_1~\!\! l_{n+1}~\!\! l_{3}~\!\! \cdots~\!\! l_{n}}
(H_{n-1})_{l_{n+1}~\!\! l_{2}~\!\! \cdots~\!\! l_{n}}
\nonumber \\
&~~~= A(t)_{l_1~\!\! l_2~\!\! \cdots~\!\! l_{n}}
(H_{n-2})_{l_1~\!\! \cdots~\!\! l_{n-2}~\!\! l_{n}~\!\! l_{n}}
(H_{n-3})_{l_1~\!\! \cdots~\!\! l_{n-3}~\!\! l_{n}~\!\! l_{n-1}~\!\! l_{n}}
\nonumber \\
&~~~~~~ \times \cdots \times
(H_{2})_{l_1~\!\! l_{2}~\!\! l_{n}~\!\! l_{4}~\!\! \cdots~\!\! l_{n}}
(H_{1})_{l_1~\!\! l_{n}~\!\! l_{3}~\!\! \cdots~\!\! l_{n}}
(H_{n-1})_{l_{n}~\!\! l_{2}~\!\! \cdots~\!\! l_{n}}
\nonumber \\
&~~~= A(t)_{l_1~\!\! l_2~\!\! \cdots~\!\! l_{n}}
\left\{\prod_{r=1}^{n-1} (h_r)_{l_r~\!\! l_n}
+ \sum_{k=2}^{n-2}\prod_{r=1}^{n-k-1} (h_r)_{l_{r+k}~\!\! l_n}
\prod_{r'=n-k}^{n-1} (h_{r'})_{l_{r'+k+1}~\!\! l_n} + \cdots\right\},
\label{A-AHs}
\end{align}
where we use the relation:
\begin{align}
(H_{k})_{l_1~\!\! \cdots~\!\! l_{k}~\!\! l_{n}~\!\! l_{k+2}~\!\! \cdots~\!\! l_{n}}
&= \sum_{u(\ne k+1, n)} (h_{k})_{l_u~\!\! l_n}
\nonumber \\
&= (h_{k})_{l_1~\!\! l_n} + \cdots + (h_{k})_{l_{k}~\!\! l_n} 
+ (h_{k})_{l_{k+2}~\!\! l_n} + \cdots + (h_{k})_{l_{n-1}~\!\! l_n}
\label{A-Hk}
\end{align}
for $k=1, 2, \cdots, n-2$ and the relation:
\begin{align}
(H_{n-1})_{l_n~\!\! l_2~\!\! \cdots~\!\! l_{n}}
&= \sum_{u(\ne 1, n)} (h_{k})_{l_u~\!\! l_n},
\nonumber \\
&= (h_{n-1})_{l_2~\!\! l_n} + (h_{n-1})_{l_{3}~\!\! l_n} 
 + \cdots + (h_{n-1})_{l_{n-1}~\!\! l_n}.
\label{A-Hn-1}
\end{align}
In eqs.~\eqref{A-Hk} and \eqref{A-Hn-1}, 
$(l_1, \cdots, l_k, l_{k+2}, \cdots l_n)$
and $(l_2, l_3, \cdots, l_{n})$ are all different, respectively.
The ellipsis in eq.~\eqref{A-AHs} stands for terms including
both $(h_{a_k})_{l_{r}~\!\! l_{n}}$ and $(h_{a_{k'}})_{l_{r}~\!\! l_{n}}$
($a_k \ne a_{k'}$) and all of them are eliminated by the addition of the terms 
which come from the exchange of $H_{a_k}$ and $H_{a_{k'}}$.

Actually, by adding all the terms with the Hamiltonians interchanged, 
we obtain the relation:
\begin{align}
&\sum_{(a_1, a_2, \cdots, a_{n-1})}{\rm sgn(P)}
(A(t)H_{a_{n-2}}H_{a_{n-3}} 
\cdots H_{a_2} H_{a_1} H_{a_{n-1}})_{l_1~\!\! l_2~\!\! \cdots~\!\! l_n}
\nonumber \\
&~~~= \sum_{(a_1, a_2, \cdots, a_{n-1})}{\rm sgn(P)}
A(t)_{l_1~\!\! l_2~\!\! \cdots~\!\! l_{n}}
\left\{\prod_{r=1}^{n-1} (h_{a_r})_{l_r~\!\! l_n}
+ \sum_{k=2}^{n-2}\prod_{r=1}^{n-k-1} (h_{a_r})_{l_{r+k}~\!\! l_n}
\prod_{r'=n-k}^{n-1} (h_{a_{r'}})_{l_{r'+k+1}~\!\! l_n}\right\}
\nonumber \\
&~~~= \sum_{(a_1, a_2, \cdots, a_{n-1})}{\rm sgn(P)}
A(t)_{l_1~\!\! l_2~\!\! \cdots~\!\! l_{n}}
\left\{\prod_{r=1}^{n-1} (h_{a_r})_{l_r~\!\! l_n}
+ \sum_{k=2}^{n-2}(-1)^{k(n-2)}\prod_{r=1}^{n-1} (h_{a_r})_{l_r~\!\! l_n}\right\}
\nonumber \\
&~~~= \gamma \sum_{(a_1, a_2, \cdots, a_{n-1})}{\rm sgn(P)}
\prod_{r=1}^{n-1} (h_{a_r})_{l_r~\!\! l_n}~\!
A(t)_{l_1~\!\! l_2~\!\! \cdots~\!\! l_{n}},
\label{A-sumAHs}
\end{align}
where $\gamma = 1$ for an odd number $n (\ge 3)$ 
and $\gamma = n-2$ for an even number $n (\ge 4)$, and
we use that $\displaystyle{\prod_{r=1}^{n-k-1} (h_{a_r})_{l_{r+k}~\!\! l_n}
\prod_{r'=n-k}^{n-1} (h_{a_{r'}})_{l_{r'+k+1}~\!\! l_n}}$
are replaced by $\displaystyle{(-1)^{k(n-2)}\prod_{r=1}^{n-1} (h_{a_r})_{l_r~\!\! l_n}}$
after exchanging $(a_1, a_2, \cdots, a_{n-1})$ for $(a_{k+1}, a_{k+2}, \cdots, a_{k})$.

In the same way, the term $(-1)^{n-2-k}(H_{n-2} H_{n-3} \cdots H_{k-1} A(t) H_{k-2} \cdots
H_2 H_1 H_{n-1})_{l_1~\!\! l_2~\!\! \cdots~\!\! l_n}$ is calculated as
\begin{align}
&(-1)^{n-2-k} (H_{n-2} H_{n-3} \cdots H_{k-1} A(t) H_{k-2} \cdots
H_2 H_1 H_{n-1})_{l_1~\!\! l_2~\!\! \cdots~\!\! l_n}
\nonumber \\
&~~~~= (-1)^{n-k} \sum_{l_{n+1}} (H_{n-2})_{l_1~\!\! l_2~\!\! \cdots~\!\! l_{n+1}}
(H_{n-3})_{l_1~\!\! \cdots~\!\! l_{n-2}~\!\! l_{n+1}~\!\! l_{n}}
\cdots (H_{k-1})_{l_1~\!\! \cdots~\!\! l_{k}~\!\! l_{n+1}~\!\! l_{k+2}~\!\! \cdots~\!\! l_{n}}
\nonumber \\
&~~~~~~~~ \times 
A(t)_{l_1~\!\! \cdots~\!\! l_{k-1}~\!\! l_{n+1}~\!\! l_{k+1}~\!\! \cdots~\!\! l_{n}}
(H_{k-2})_{l_1~\!\! \cdots~\!\! l_{k-2}~\!\! l_{n+1}~\!\! l_{k}~\!\! \cdots~\!\! l_{n}}
 \cdots (H_{2})_{l_1~\!\! l_{2}~\!\! l_{n+1}~\!\! l_{4}~\!\! \cdots~\!\! l_{n}}
(H_{1})_{l_1~\!\! l_{n+1}~\!\! l_{3}~\!\! \cdots~\!\! l_{n}}
\nonumber \\
&~~~~~~~~ \times
(H_{n-1})_{l_{n+1}~\!\! l_{2}~\!\! \cdots~\!\! l_{n}}
\nonumber \\
&~~~~ = (-1)^{n-k} A(t)_{l_1~\!\! l_2~\!\! \cdots~\!\! l_{k-1}~\!\! 
l_{k}~\!\! l_{k+1}~\!\! \cdots~\!\! l_{n}}
(H_{n-2})_{l_1~\!\! l_2~\!\! \cdots~\!\! l_{n-1}~\!\! l_k}
(H_{n-3})_{l_1~\!\! \cdots~\!\! l_{n-2}~\!\! l_{k}~\!\! l_{n}}
\cdots (H_{k-1})_{l_1~\!\! \cdots~\!\! l_{k}~\!\! l_{k}~\!\! l_{k+2}~\!\! \cdots~\!\! l_{n}}
\nonumber \\
&~~~~~~~~  \times 
(H_{k-2})_{l_1~\!\! \cdots~\!\! l_{k-2}~\!\! l_{k}~\!\! l_{k}~\!\! \cdots~\!\! l_{n}}
 \cdots (H_{2})_{l_1~\!\! l_{2}~\!\! l_{k}~\!\! l_{4}~\!\! \cdots~\!\! l_{n}}
(H_{1})_{l_1~\!\! l_{k}~\!\! l_{3}~\!\! \cdots~\!\! l_{n}}
(H_{n-1})_{l_{k}~\!\! l_{2}~\!\! \cdots~\!\! l_{n}}
\nonumber \\
&~~~~= (-1)^{n-k} A(t)_{l_1~\!\! l_2~\!\! \cdots~\!\! l_{n}}
\left\{\prod_{r=1}^{k-1} (h_r)_{l_{r}~\!\! l_k}
\prod_{r'=k}^{n-1} (h_{r'})_{l_{r'+1}~\!\! l_k}\right.
\nonumber \\
&~~~~~~~~+ \sum_{m=2}^{k-3}\prod_{r=1}^{k-1-m} (h_r)_{l_{r+m}~\!\! l_k}
\prod_{r'=k-m}^{n-1} (h_{r'})_{l_{r'+m+1}~\!\! l_k}
+ (h_{1})_{l_{k-1}~\!\! l_k}\prod_{r=2}^{n-1} (h_r)_{l_{r+k-1}~\!\! l_k}
+ \prod_{r=1}^{n-1} (h_r)_{l_{r+k}~\!\! l_k}
\nonumber \\
&~~~~~~~~ \left.
+ \sum_{m={k+1}}^{n-1}\prod_{r=1}^{k-1-m} (h_r)_{l_{r+m}~\!\! l_k}
\prod_{r'=k-m}^{n-1} (h_{r'})_{l_{r'+m+1}~\!\! l_k} + \cdots\right\},
\label{A-AHs-k}
\end{align}
where we use the relation:
\begin{align}
(H_{j})_{l_1~\!\! \cdots~\!\! l_{j}~\!\! l_{k}~\!\! l_{j+2}~\!\! 
\cdots~\!\! l_{k-1}~\!\! l_{k}~\!\! l_{k+1}~\!\! \cdots~\!\! l_{n}}
&= \sum_{u(\ne j+1, k)} (h_{j})_{l_u~\!\! l_k}
\nonumber \\
&= (h_{j})_{l_1~\!\! l_k} + \cdots + (h_{j})_{l_{j}~\!\! l_k} 
+ (h_{j})_{l_{j+2}~\!\! l_k} 
\nonumber \\
& ~~~~ + \cdots + (h_{j})_{l_{k-1}~\!\! l_k} 
+ (h_{j})_{l_{k+1}~\!\! l_k} + \cdots + (h_{j})_{l_{n}~\!\! l_k}
\label{A-Hj<}
\end{align}
for $1 \le j \le k-3$, 
\begin{align}
(H_{k-2})_{l_1~\!\! \cdots~\!\! l_{k}~\!\! l_{k}~\!\! l_{k+1}~\!\! \cdots~\!\! l_{n}}
&= \sum_{u(\ne k-1, k)} (h_{k-2})_{l_u~\!\! l_k}
\nonumber \\
&= (h_{k-2})_{l_1~\!\! l_k} + \cdots + (h_{k-2})_{l_{k-2}~\!\! l_k} 
+ (h_{k-2})_{l_{k+1}~\!\! l_k} + \cdots + (h_{k-2})_{l_{n}~\!\! l_k},
\label{A-Hk-2}\\
(H_{k-1})_{l_1~\!\! \cdots~\!\! l_{k}~\!\! l_{k}~\!\! l_{k+2}~\!\! \cdots~\!\! l_{n}}
&= \sum_{u(\ne k, k+1)} (h_{k-1})_{l_u~\!\! l_k}
\nonumber \\
&= (h_{k-1})_{l_1~\!\! l_k} + \cdots + (h_{k-1})_{l_{k-1}~\!\! l_k} 
+ (h_{k-1})_{l_{k+2}~\!\! l_k} + \cdots + (h_{k-1})_{l_{n}~\!\! l_k}
\label{A-Hk-1}
\end{align}
and
\begin{align}
(H_{j})_{l_1~\!\! \cdots~\!\! l_{k-1}~\!\! l_{k}~\!\! l_{k+1}~\!\! 
\cdots~\!\! l_{j+1}~\!\! l_{k}~\!\! l_{j+3}~\!\! \cdots~\!\! l_{n}}
&= \sum_{u(\ne k, j+2)} (h_{j})_{l_u~\!\! l_k}
\nonumber \\
&= (h_{j})_{l_1~\!\! l_k} + \cdots + (h_{j})_{l_{k-1}~\!\! l_k} 
+ (h_{j})_{l_{k+1}~\!\! l_k} 
\nonumber \\
& ~~~~ + \cdots + (h_{j})_{l_{j+1}~\!\! l_k} 
+ (h_{j})_{l_{j+3}~\!\! l_k} + \cdots + (h_{j})_{l_{n}~\!\! l_k}
\label{A-Hj>}
\end{align}
for $k \le j \le n-1$.
In eqs.~\eqref{A-Hj<}, \eqref{A-Hk-2}, \eqref{A-Hk-1} and \eqref{A-Hj>}, 
the $n-1$ indices
$(l_1, \cdots, l_j, l_{j+2}, \cdots l_n)$,
$(l_1, \cdots, l_{k-2}, l_{k}, \cdots l_n)$,
$(l_1, \cdots, l_{k}, l_{k+2}, \cdots l_n)$,
and $(l_1, \cdots, l_{j+1}, l_{j+3}, \cdots l_n)$ are all different, respectively.
The ellipsis in eq.~\eqref{A-AHs-k} stands for terms including
both $(h_{a_k})_{l_{r}~\!\! l_{n}}$ and $(h_{a_{k'}})_{l_{r}~\!\! l_{n}}$
($a_k \ne a_{k'}$) and all of them are eliminated by adding the terms which stem from
the exchange of $H_{a_k}$ and $H_{a_{k'}}$.

By adding all the terms with the Hamiltonians interchanged, we obtain the relation:
\begin{align}
&(-1)^{n-k} \sum_{(a_1, a_2, \cdots, a_{n-1})}{\rm sgn(P)}
(H_{a_{n-2}} H_{a_{n-3}} \cdots H_{a_{k-1}} A(t) H_{a_{k-2}} \cdots
H_{a_2} H_{a_1} H_{a_{n-1}})_{l_1~\!\! l_2~\!\! \cdots~\!\! l_n}
\nonumber \\
&~~~= (-1)^{n-k} \sum_{(a_1, a_2, \cdots, a_{n-1})}{\rm sgn(P)}
A(t)_{l_1~\!\! l_2~\!\! \cdots~\!\! l_{n}}
\left\{\prod_{r=1}^{k-1} (h_{a_r})_{l_{r}~\!\! l_k}
\prod_{r'=k}^{n-1} (h_{a_{r'}})_{l_{r'+1}~\!\! l_k}\right.
\nonumber \\
&~~~~~~~ 
+ \sum_{m=2}^{k-3}\prod_{r=1}^{k-1-m} (h_{a_r})_{l_{r+m}~\!\! l_k}
\prod_{r'=k-m}^{n-1} (h_{a_{r'}})_{l_{r'+m+1}~\!\! l_k}
+ (h_{a_1})_{l_{k-1}~\!\! l_k}\prod_{r=2}^{n-1} (h_{a_r})_{l_{r+k-1}~\!\! l_k}
+ \prod_{r=1}^{n-1} (h_{a_r})_{l_{r+k}~\!\! l_k}
\nonumber \\
&~~~~~~~ \left. + \sum_{m={k+1}}^{n-1}\prod_{r=1}^{k-1-m} (h_{a_r})_{l_{r+m-1}~\!\! l_k}
\prod_{r'=k-m}^{n-1} (h_{a_{r'}})_{l_{r'+m+1}~\!\! l_k}
+ \cdots\right\}
\nonumber \\
&~~~= (-1)^{n-k} \sum_{(a_1, a_2, \cdots, a_{n-1})}{\rm sgn(P)}
A(t)_{l_1~\!\! l_2~\!\! \cdots~\!\! l_{n}}
\left\{\prod_{r=1}^{k-1} (h_{a_r})_{l_{r}~\!\! l_k}
\prod_{r'=k}^{n-1} (h_{a_{r'}})_{l_{r'+1}~\!\! l_k}\right.
\nonumber \\
&~~~~~~~ \left.
+ \sum_{m=2}^{n-2}(-1)^{m(n-2)}
\prod_{r=1}^{k-1} (h_{a_r})_{l_{r}~\!\! l_k}
\prod_{r'=k}^{n-1} (h_{a_{r'}})_{l_{r'+1}~\!\! l_k}\right\}
\nonumber \\
&~~~= \gamma (-1)^{n-k} \sum_{(a_1, a_2, \cdots, a_{n-1})}{\rm sgn(P)}
\prod_{r=1}^{k-1} (h_{a_r})_{l_{r}~\!\! l_k}
\prod_{r'=k}^{n-1} (h_{a_{r'}})_{l_{r'+1}~\!\! l_k}~\!
A(t)_{l_1~\!\! l_2~\!\! \cdots~\!\! l_{n}},
\label{A-sumAHs-k}
\end{align}
where $\gamma = 1$ for an odd number $n (\ge 3)$ 
and $\gamma = n-2$ for an even number $n (\ge 4)$, and
we use that the factors such as
\begin{align} 
&\sum_{m=2}^{k-3}\prod_{r=1}^{k-1-m} (h_{a_r})_{l_{r+m}~\!\! l_k}
\prod_{r'=k-m}^{n-1} (h_{a_{r'}})_{l_{r'+m+1}~\!\! l_k},~~
(h_{a_1})_{l_{k-1}~\!\! l_k}\prod_{r=2}^{n-1} (h_{a_r})_{l_{r+k-1}~\!\! l_k}, 
\nonumber \\
&\prod_{r=1}^{n-1} (h_{a_r})_{l_{r+k}~\!\! l_k},~~
\sum_{m={k+1}}^{n-1}\prod_{r=1}^{k-1-m} (h_{a_r})_{l_{r+m-1}~\!\! l_k}
\prod_{r'=k-m}^{n-1} (h_{a_{r'}})_{l_{r'+m+1}~\!\! l_k}
\label{A:factors}
\end{align}
are replaced by 
\begin{align} 
&\sum_{m=2}^{k-3}(-1)^{m(n-2)}\prod_{r=1}^{k-1} (h_{a_r})_{l_{r}~\!\! l_k}
\prod_{r'=k}^{n-1} (h_{a_{r'}})_{l_{r'+1}~\!\! l_k},~~
(-1)^{(k-2)(n-2)} \prod_{r=1}^{k-1} (h_{a_r})_{l_{r}~\!\! l_k}
\prod_{r'=k}^{n-1} (h_{a_{r'}})_{l_{r'+1}~\!\! l_k},
\nonumber \\
&(-1)^{(k-1)(n-2)}\prod_{r=1}^{k-1} (h_{a_r})_{l_{r}~\!\! l_k}
\prod_{r'=k}^{n-1} (h_{a_{r'}})_{l_{r'+1}~\!\! l_k},~~
\sum_{m={k+1}}^{n-1}(-1)^{(m-1)(n-2)}\prod_{r=1}^{k-1} (h_{a_r})_{l_{r}~\!\! l_k}
\prod_{r'=k}^{n-1} (h_{a_{r'}})_{l_{r'+1}~\!\! l_k},
\label{A:factors-2}
\end{align}
respectively, after exchanging $(a_1, a_2, \cdots, a_{n-1})$ for suitable ones.
We notice that eq.~\eqref{A-sumAHs} corresponds to the case with $k=n$
in eq.~\eqref{A-sumAHs-k}.

The left-hand side of eq.~\eqref{A-gH-eq-n}
becomes as
\begin{eqnarray}
\frac{d}{dt} A_{l_1~\!\! l_2~\!\! \cdots~\!\! l_n}(t) 
= 2\pi i (\nu_{n})_{l_1~\!\! l_2~\!\! \cdots~\!\! l_n}
A_{l_1~\!\! l_2~\!\! \cdots~\!\! l_n}(t).
\label{A-gH-eq-n-lhs}
\end{eqnarray}
On the other hand, the right-hand side of eq.~\eqref{A-gH-eq-n}
becomes as
\begin{align}
\frac{1}{i \hbar} 
\left[A(t), H_1, H_2, \cdots, H_{n-1}\right]_{l_1~\!\! l_2~\!\! \cdots~\!\! l_n}
&=-\frac{2\pi i}{h}(-1)^{\frac{(n-2)(n-3)}{2}}
\left[A(t), H_{n-2}, \cdots, H_2, H_1, H_{n-1}\right]_{l_1~\!\! l_2~\!\! \cdots~\!\! l_n}
\nonumber \\
&=\frac{2\pi i}{h}(-1)^{\frac{(n-2)(n-3)}{2}+1}
 \gamma \sum_{k=1}^n (-1)^{n-k} \sum_{(a_1, a_2, \cdots, a_{n-1})}{\rm sgn(P)}
\nonumber \\
&~~~ \times \prod_{r=1}^{k-1} (h_{a_r})_{l_{r}~\!\! l_k}
\prod_{r'=k}^{n-1} (h_{a_{r'}})_{l_{r'+1}~\!\! l_k}~\!
A(t)_{l_1~\!\! l_2~\!\! \cdots~\!\! l_{n}},
\label{A-gH-eq-n-rhs}
\end{align}
where we use eqs.~\eqref{A-gH-rhs} and \eqref{A-sumAHs-k}.
From eqs.~\eqref{A-gH-eq-n-lhs} and \eqref{A-gH-eq-n-rhs},
eq.~\eqref{A-nu-N-qu-n-h} is obtained and hence it is shown that
$A(t)_{l_1~\!\! l_2~\!\! \cdots~\!\! l_{n}}$ 
given eqs.~\eqref{A-A(t)} and \eqref{A-nu-N-qu-n-h}
satisfies eq.~\eqref{A-gH-eq-n}.

Incidentally, we point out that, 
if the generalized Heisenberg equation is defined by 
\begin{eqnarray}
\frac{d}{dt} A_{l_1~\!\! l_2~\!\! \cdots~\!\! l_n}(t) 
= \frac{1}{i\hbar}
\left[A(t), H_{n-2}, \cdots, H_2, H_1, H_{n-1}\right]_{l_1~\!\! l_2~\!\! \cdots~\!\! l_n},
\label{A-gH-eq-n-redef}
\end{eqnarray}
the formula of $(\nu_n)_{l_1~\!\! l_2~\!\! \cdots l_n}$ takes the form:
\begin{align}
(\nu_n)_{l_1~\!\! l_2~\!\! \cdots l_n} 
= -\frac{\gamma}{h}
\sum_{k=1}^{n} (-1)^{n-k} \sum_{(a_1, a_2, \cdots, a_{n-1})}{\rm sgn(P)} 
\prod_{r=1}^{k-1}(h_{a_r})_{l_r~\!\! l_k}\prod_{r'=k}^{n-1}(h_{a_{r'}})_{l_{r'+1}~\!\! l_k}
\label{A-nu-N-qu-n-h-redef}
\end{align}
due to the disappearance of the factor $(-1)^{\frac{(n-2)(n-3)}{2}}$,
as seen from eqs.~\eqref{A-nu-N-qu-n-h}, \eqref{A-gH-eq-n} and \eqref{A-gH-rhs}.

\section{Reduction to the Heisenberg dynamics}

In the subsection 3.1, we have seen how Nambu dynamics is reduced to Hamiltonian dynamics.
In concrete, the Nambu equation turns out to be 
Hamilton's canonical equations of motion by reducing the phase space.

In this appendix, we show that the generalized Heisenberg equation can be reduced to 
Heisenberg's equation of motion by taking a specific form of Hamiltonians.
The starting point is the extension of Heisenberg's equation of motion:
\begin{eqnarray}
\frac{d}{dt} A^{\alpha}_{l~\!\! m~\!\! n}(t) 
= \frac{1}{i \hbar} \left[A^{\alpha}(t), H_1, H_2\right]_{l~\!\! m~\!\! n},
\label{gH-eq-B}
\end{eqnarray}
where $A^{\alpha}_{l~\!\! m~\!\! n}(t)$ ($l, m, n = 0, 1, 2 \cdots, N$, $\alpha =1,2,3$) 
are the components of 
$(N+1) \times (N+1) \times (N+1)$ generalized matrices $A^{\alpha}=(X, Y, Z)$,
$H_1$ and $H_2$ are Hamiltonians given by $(N+1) \times (N+1) \times (N+1)$ generalized matrices.
We note that $N$ can be infinity depending on a system.

In general, $H_1$ depends on some variables of $A^{\alpha}$
and it takes specific values depending on the solution.
Let us suppose that the following solution is given by
\begin{eqnarray}
&~& X_{l~\!\! m~\!\! n}(t) 
= \begin{cases}
X_{l~\!\! m~\!\! n}(0)e^{2\pi i\nu_{m~\!\! n}t} & (l=0, m, n \ne 0)\\
0 & ({\rm otherwise})
\end{cases},
\label{X-sol}\\
&~& Y_{l~\!\! m~\!\! n}(t) 
= \begin{cases}
Y_{l~\!\! m~\!\! n}(0)e^{2\pi i\nu_{m~\!\! n}t} & (l=0, m, n \ne 0)\\
0 & ({\rm otherwise})
\end{cases}
\label{Y-sol}
\end{eqnarray}
and $H_1$ takes the values such as
\begin{eqnarray}
(H_1)_{l~\!\! m~\!\! n}
= \begin{cases}
E_m \delta_{m~\!\! n} & (l=0, m, n \ne 0)\\
0 & ({\rm otherwise}),
\end{cases}
\label{H1-sol}
\end{eqnarray}
where $X_{l~\!\! n~\!\! m}(0)=\overline{X_{l~\!\! m~\!\! n}(0)}$,
$Y_{l~\!\! n~\!\! m}(0)=\overline{Y_{l~\!\! m~\!\! n}(0)}$
and $h\nu_{m~\!\! n} = E_m - E_n$.

When $(H_2)_{l~\!\! m~\!\! n}=1$ for $l, m, n \ne 0$,
the equation \eqref{gH-eq-B} can be rewritten
in the form of Heisenberg's equation of motion:
\begin{eqnarray}
\frac{d}{dt} A^{\alpha}_{l~\!\! m~\!\! n}(t) 
= \frac{1}{i \hbar} \left[A^{\alpha}(t), H\right]'_{l~\!\! m~\!\! n},
\label{H-eq-B}
\end{eqnarray}
where $H=H_1$ and $\left[A^{\alpha}(t), H\right]'_{l~\!\! m~\!\! n}$ is defined by
\begin{eqnarray}
\left[A^{\alpha}(t), H\right]'_{l~\!\! m~\!\! n}
\equiv \sum_{k =1}^{N}\left(A^{\alpha}(t)_{l~\!\! m~\!\! k}H_{l~\!\! k~\!\! n}
- H_{l~\!\! m~\!\! k}A^{\alpha}(t)_{l~\!\! k~\!\! n}\right).
\label{commutator-B}
\end{eqnarray}
Furthermore, the commutation relation  
$[X, Y, Z]_{l~\!\! m~\!\! n} = i\hbar I_{l~\!\! m~\!\! n}$ can be reduced to the usual one
$[X, Y, Z]_{0~\!\! m~\!\! n} = [X, Y]'_{0~\!\! m~\!\! n} =i\hbar \delta_{m~\! n}$
for $m, n \ne 0$, if we choose $Z=H_2$.

In this way, we find that Heisenberg's matrix mechanics can be described by 
the generalized matrices.

\end{document}